\documentclass[12pt,preprint]{aastex}
\usepackage{epsfig}

\def\lesssim{\mathrel{\hbox{\rlap{\hbox{\lower4pt\hbox{$\sim$}}}\hbox{$<$}}}}
\def\gtrsim{\mathrel{\hbox{\rlap{\hbox{\lower4pt\hbox{$\sim$}}}\hbox{$>$}}}}

\received{2003 May 19}
\begin{document}

\title{Collapse of Magnetized Singular Isothermal Toroids: II.
Rotation and Magnetic Braking}

\author{Anthony Allen}
\affil{Institute of Astronomy and Astrophysics, Academia Sinica,
PO BOX 23-141, Taipei 106, Taiwan, R.O.C.} 

\author{Zhi-Yun Li}
\affil{Department of Astronomy, University of Virginia, Charlottesville, 
VA 22903}

\author{Frank H. Shu}
\affil{National Tsing Hua University, 101, Section 2 Kuang Fu Road, 
Hsinchu, Taiwan 300, R.O.C.}

\begin{abstract}
We study numerically the collapse of rotating, magnetized molecular cloud 
cores, focusing on rotation and magnetic braking during the main accretion 
phase of isolated star formation. Motivated by previous numerical work and 
analytic considerations, we idealize the pre-collapse core as a magnetized singular 
isothermal toroid, with a constant rotational speed everywhere. The 
collapse starts from the center, and propagates outwards in an inside-out 
fashion, satisfying exact self-similarity in space and time.
For rotation rates and field strengths typical of dense low-mass
cores, the main feature remains the flattening of the mass distribution along 
field lines -- the formation of a pseudodisk, as in the nonrotating cases. 
The density distribution of the pseudodisk is little affected by rotation. 
On the other hand,
the rotation rate is strongly modified by pseudodisk formation. Most of the 
centrally accreted material reaches the 
vicinity of the protostar through the pseudodisk. The specific angular 
momentum can be greatly reduced on the way, by an order of magnitude or 
more, even when the pre-collapse field strength is substantially below 
the critical value for dominant cloud support. The efficient magnetic braking is due to the pinched 
geometry of the magnetic field in the pseudodisk, which strengthens the 
magnetic field and lengthens the level arm for braking. Both effects enhance 
the magnetic transport of angular momentum from inside to outside.
The excess angular momentum is carried away in a low-speed 
outflow that has, despite claims made by other workers, little in
common with observed bipolar molecular outflows.  We discuss the
implications of our calculations for the formation of true disks that are 
supported against gravity by rotation.  

\end{abstract}

\keywords{accretion --- ISM: clouds --- magnetohydrodynamics --- stars: 
formation}

\section{Introduction}

Rotation is observed ubiquitously in star-forming cores. On the scale 
of $\sim$0.1~pc, the detected rotation rate is small, with a typical
value corresponding to a ratio of rotational to gravitational energy 
of $0.02$ (Goodman et al. 1993). The rotation rate is much lower than 
expected if the core is condensed out of a low density background 
medium with angular momentum conserved. The slow core rotation is 
thought to be due to magnetic braking (Mestel 1985; Mouschovias \& 
Ciolek 1999), which is particularly efficient during the long, 
magnetically subcritical phase of cloud evolution driven by ambipolar 
diffusion (Basu \& Mouschovias 1994).  This assumes that the clouds 
are strongly magnetized to begin with, as envisioned in the standard
picture of isolated low-mass star formation (Shu, Adams \& Lizano
1987). Once a magnetically supercritical core develops, it begins 
to collapse dynamically. It is often assumed that magnetic braking 
then becomes inefficient, with most of the angular momentum of the 
core material carried by the collapsing flow into the central 
star-plus-disk or binary system. However, the typical value of 
specific angular momentum inferred for dense cores is an order of 
magnitude higher than the typical binary value, at least in the Taurus 
cloud complex (see Fig.~8 of Simon et al. 1995 and discussion 
in \S~8.5 of Myers 1995). This suggests that a large fraction 
of the core angular momentum may still be removed {\it during collapse}.
The problem of angular momentum redistribution in a collapsing 
magnetized core thus warrants a closer examination. 

The efficiency of magnetic braking depends sensitively on the magnetic
field geometry (Mestel 1985). The exact field geometry of a magnetized 
core on the verge of collapse is difficult to determine without 
following the cloud evolution in three dimensions. Simplified numerical 
calculations  of cloud evolution driven by ambipolar diffusion have 
shown, however, that by the time a central density cusp is formed the 
mass-to-flux ratio is approximately spatially constant in the 
supercritical core, and the (volume) density decreases with radius 
roughly as a power-law of $r^{-2}$ (Lizano \& Shu 1989; Basu \& 
Mouschovias 1994). This self-similar behavior motivated Li \& Shu 
(1996) to idealize the end state of the ambipolar diffusion-driven core 
formation as a scale-free configuration supported in part by a large 
scale magnetic field and in part by the thermal pressure gradient,
with a fixed mass-to-flux ratio and an exact $r^{-2}$ density
profile. Such a configuration turns out to be a singular isothermal 
toroid, for which the  magnetic field geometry can be computed exactly
(see also Baureis, Ebert, \& Schmitz 1989).  The field geometry 
is distorted by fluid motions during the subsequent inside-out collapse.

In a companion paper to this journal (Allen, Shu \& Li 2003; hereafter 
Paper I), we have studied the collapse of magnetized, singular isothermal 
toroids without rotation. The main features of the collapse solution 
are governed by the anisotropy in the magnetic forces, which induces
in the toroid an evacuated region near the axis, where the plasma 
$\beta$ parameter (the ratio of thermal to magnetic energy density) 
drops below unity (see Fig.~4a of Paper I). As the core collapses from 
inside out, matter in the magnetically dominated region slides along 
the field lines towards the equatorial region, creating a dense, 
flattened structure -- a pseudodisk (Galli \& Shu 1993a,b). The 
material in the pseudodisk collapses radially towards the central 
point mass in 
a magnetically diluted free-fall, dragging the footpoints of field 
lines along with it (see Fig.~4b of Paper I). A highly pinched 
field configuration (``split monopole") is produced,
with important implications for the magnetic 
transport of angular momentum from inside to outside.
In the presence of rotation, the collapse would lead to a 
spinup of the pseudodisk, which is linked magnetically to a more
slowly rotating 
envelope of infalling and nearly static material.  We will show 
in this paper that the twisting of magnetic field lines actually 
drives a slow 
outflow during the core collapse, which can remove much of the angular 
momentum remaining in the infalling pseudodisk.  This removal poses a
substantial brake against the eventual formation of a centrifugally
supported disk unless the trapped magnetic field is lost in the realistic
situation by nonideal MHD effects.

As the first step toward the dynamical calculations, we will
generalize the magnetized, singular isothermal toroid 
solutions of Li \& Shu (1996) to include rotation in \S~\ref{initial}. 
These magnetized rotating toroids are then induced to collapse from 
inside out in \S~\ref{collapse}, and the collapse is followed numerically 
in a manner similar to that described in Paper I. The numerical solutions 
are discussed in \S~\ref{discussion}, with an emphasis on the efficiency 
of magnetic braking during core collapse and the implications for the 
difficulty of forming rotationally supported disks unless significant 
flux loss appears in the realistic problem at small scales, as is 
expected theoretically for high-density situations (e.g., Nishi, Nakano 
\& Umebayashi 1991; Desch \& Mouschovias 2001).
We end with a brief summary and conclusion in \S~\ref{conclusion}.

\section{Magnetized, Rotating Singular Isothermal Toroids}
\label{initial}

\subsection{Governing Equations} 
\label{general}

We consider a rotating, scale-free (self-similar), self-gravitating
cloud that is magnetized by a purely poloidal field and in mechanical
force balance. The cloud is specified by the mass density $\rho$, the 
magnetic flux function $\Phi$ which determines the poloidal field 
through $2\pi {\bf B}=\nabla\times[(\Phi/r\sin\theta){\bf e}_\varphi]$, 
and the rotational speed $u_\varphi$, in a spherical polar coordinate 
system $(r,\theta,\varphi)$. Under the assumption of axisymmetry, the 
radial and angular dependences of these cloud quantities can be 
separated, on dimensional grounds, into 
\begin{equation}
\rho(r,\theta)={a^2\over 2\pi G r^2} R(\theta),
\label{e1}
\end{equation}
\begin{equation}
\Phi(r,\theta)={4\pi a^2 r\over G^{1/2}}\phi(\theta),
\label{e2}
\end{equation}
\begin{equation}
u_\varphi(r,\theta)=a\ v(\theta),
\label{e3}
\end{equation}
where $a$ is the isothermal sound speed, which we take to be a constant,
and the functions $R(\theta)$, $\phi(\theta)$ and $v(\theta)$ describe
the angular distributions of the density, magnetic flux, and rotational
speed. 

In order to be in a mechanical force balance, the cloud must satisfy the 
momentum equation, 
\begin{equation}
\rho\left[ \nabla\left({u_\varphi^2\over 2}
	\right) + (\nabla\times{\bf u})\times {\bf u}\right]=
	-\rho \nabla V - \nabla P + {1\over 4\pi} (\nabla\times {\bf B})
	\times {\bf B},
\label{e4}
\end{equation}
where $P=\rho a^2$ is the thermal pressure, and the velocity ${\bf u}=
u_\varphi {\bf e}_\varphi$ has only a toroidal component. The gravitational 
potential $V$ is related to the density through Poisson's equation
\begin{equation}
\nabla^2V=4\pi G \rho.
\label{e5}
\end{equation}
It can be decomposed into a form (Toomre 1982)  
\begin{equation}
V(r,\theta)=2 a^2 (1+H_0)[\ln r + h(\theta)],
\label{e6}
\end{equation}
where $H_0$ is the fractional over-density of the (rotating) 
magnetized configuration over the singular isothermal sphere that 
was introduced by Li \& Shu (1996). This overdensity arises 
from a combination of 
magnetization and rotation. The angular function 
$h(\theta)$ is equivalent to the function $P(\theta)$ of Toomre (1982;
see his equation~[7]). 

Substituting equations (\ref{e1}), (\ref{e2}), (\ref{e3}) and (\ref{e6}) 
into equations (\ref{e4}) and (\ref{e5}) and eliminating the function 
$h(\theta)$ yield two coupled second-order ordinary differential 
equations (ODEs) for the angular distributions of density, $R(\theta)$, 
and magnetic flux, $\phi(\theta)$,
\begin{equation}
{d\over d\theta}\left[\sin\theta\left( -{1\over R}
{d R\over d\theta} -{v^2-2H_0\over \phi}{d\phi\over d\theta} +
{v^2\cos\theta\over \sin\theta}\right)\right] = 2(R-1-H_0) \sin\theta,
\label{e7}
\end{equation}
\begin{equation}
{d\over d\theta}\left({1\over \sin\theta}{d\phi\over d\theta}\right)
={R\sin\theta\over 2\phi}(v^2-2H_0),
\label{e8}
\end{equation}
where the angular distribution of rotational speed $v(\theta)$ remains 
to be determined. In the nonrotating limit that $v(\theta)=0$, these
two coupled ODEs reduce to equations (12) and (13) of Li \& Shu (1996). 

The eliminated function $h(\theta)$ that appears in the expression
for gravitational potential, equation (\ref{e6}), is related to other 
functions by 
\begin{equation}
{d h\over d\theta}={1\over 2(1+H_0)}\left[ -{1\over R}
{d R\over d\theta} -{v^2-2H_0\over \phi}{d\phi\over 
d\theta} + {v^2\cos\theta\over \sin\theta}\right].
\label{e9}
\end{equation}

\subsection{Non-Magnetic Limit}
\label{non-magnetic}

In the nonmagnetic limit $\phi(\theta)\to 0$, equation~(\ref{e8})
demands that $v(\theta)=\sqrt{2H_0}$, i.e., constant rotational
speed everywhere. It is easy to verify the remaining two governing
equations, (\ref{e7}) and (\ref{e9}), reduce to equations (15)
and (16) of Toomre (1982; see also Hayashi, Narita \& Miyama 1982), 
with slightly different notations. In
this purely rotating case, Toomre (1982) gave an exact solution 
for the angular distribution of density, which in our notation 
becomes
\begin{equation}
R(\theta)=(1+H_0)^2 \cosh^2\xi\ {\rm sech}^2[(1+H_0)\xi],
\label{e10}
\end{equation}
where the new variable $\xi$ is related to the polar angle $\theta$ 
through $\tanh\xi=\cos\theta$. Equation~(\ref{e10}) describes the 
so-called ``Toomre-Hayashi toroid''. 

\subsection{Generalized Toroid Solutions}

In the general case, both the flux function $\phi(\theta)$ and the
rotation speed $v(\theta)$ are nonzero. This creates a potential
problem, because fluid parcels along a given field line 
will generally wrap up the field, introducing a toroidal component
that redistributes the fluid angular momentum,
unless the angular speed $\Omega(r,\theta)$
is constant along the field line (i.e., matter and 
magnetic field corotate).  The result follows from field freezing, i.e., the
induction equation of ideal MHD, which we have yet to consider, coupled with
the condition $\nabla\cdot {\bf B} = 0$. Applied to an axisymmetric 
configuration of poloidal field lines, the presence of axial rotation, 
$u_\varphi \equiv r\Omega \sin \theta$,
will lead to a generation of a toroidal component at a rate: 
\begin{equation}
{\partial B_\varphi \over\partial t} = r\sin \theta \left[ B_r{\partial \Omega
\over \partial r}+{B_\theta\over r}{\partial \Omega\over \partial 
\theta}\right].
\label{induceq}
\end{equation}
For there to be no toroidal field generation or change,
the right-hand side has to be
zero, i.e., ${\bf B}\cdot \nabla \Omega = 0$, which is the
requirement of isorotation on field lines (Ferraro 1937).
In a self-similar toroid, isorotation requires 
\begin{equation}
v(\theta)={C \sin\theta\over \phi(\theta)},
\label{e11}
\end{equation}
where $C$ is a constant. 

Unfortunately, self-similar magnetic field lines providing nontrivial 
cloud support extend infinitely far in 
cylindrical radius $r\sin\theta$. If $\Omega$ were to remain a nonzero
constant as $r\sin\theta \rightarrow \infty$
on a field line, then the linear rotational 
speed $u_\varphi =r\Omega \sin\theta$
would increase without bound. The resulting
divergent centrifugal force cannot be balanced by any combination
of the other self-similar forces in the problem. 
Differential rotation along a field line
is therefore unavoidable if the unbounded cloud is to start 
in mechanical force balance everywhere.  Alternatively, we may say 
that if $\Omega$
has a nonzero constant value on a field line, then the quantity
$a/\Omega$ would yield an intrinsic length scale, and we would 
have to give up the simplifying assumption of self-similarity. We 
choose to retain exact self-similarity and forego instead isorotation
on field lines.

Differential rotation along a field line, ${\bf B}\cdot \nabla \Omega 
\neq 0$ at the initial (pivotal) instant $t=0$, extracts a price, 
and that price
is an instantaneous wrapping of the magnetic field for $t > 0$ via 
equation (\ref{induceq}).
The spontaneous generation of toroidal field is an inevitable
consequence of differential rotation in the problem, so
one might wonder why we do not include a systematic toroidal field at 
the outset
in the pivotal state.  We do not do so for three reasons. First, it
would introduce arbitrary extra parameters into a problem that is
already sufficiently complex. Second, axisymmetric
models that contain toroidal fields which 
extend infinitely in the axial direction (e.g., Fiege \& Pudritz 2000) 
generally
have undesirable unclosed systems of currents that lead to large-scale
separations of electric charge over time. Third, observations
of millimeter-wave and far infrared polarization patterns 
do not indicate substantive toroidal fields in the noncollapsing 
portions of molecular and dark clouds (see, e.g.,
Ward-Thompson et al. 2000 for the case of starless core L1544).

In our problem, the spatially limited generation 
of a finite level of $B_\varphi$ slowly throws the cloud out of mechanical 
equilibrium
in the poloidal directions.  We regard this feature of the solutions to be
less important than the fast poloidal motions that are generated because the
pivotal state is unstable to inside-out dynamical collapse.  Thus,
we shall focus our attention for $t>0$ on the magnetic torques exerted 
on exterior material connected to field lines dragged into a rotating, 
collapsing pseudodisk. We shall find the back reaction of those torques 
on the infalling material to be a formidable magnetic brake working 
against the formation of centrifugally supported disks.

A simple way to introduce the necessary degree of differential rotation
into the equilibrium toroid is to multiply the expression for
$v(\theta)$ in equation~(\ref{e11}) by an extra factor of $\sin(\theta)$, 
so that
\begin{equation}
v(\theta)={C \sin^2\theta\over \phi(\theta)}.
\label{e12}
\end{equation}
We have checked by direct computation that models
satisfying equation (\ref{e12}) have finite rotation speeds everywhere,
with a moderate angular variation. The above 
choice is arbitrary, however, and a simpler, esthetically more appealing 
alternative is a constant rotational speed $v$ independent of $\theta$.
This choice for the pivotal state is suggested by the
pre-pivotal evolutionary calculations of Basu \& Mouschovias (1994) including
both ambipolar diffusion and rotation. The choice also has
the advantage of describing the Toomre-Hayashi toroid 
in the nonmagnetic limit.  Collapse calculations computed
with pivotal states satisfying equation (\ref{e12}) yield results
that are qualitatively similar to the cases with $v=$ const.  In addition, 
self-gravitating isothermal configurations with flat rotation curves have 
a long respected tradition in observational astronomy and theoretical 
astrophysics, with generalizations even possible into the relativistic
regime (Cai \& Shu 2002, 2003) and for nonaxisymmetric equilibria (Syer 
\& Tremaine 1996, Galli et al. 2001).

With $v$ = const $\equiv v_0$, we can now cast the two coupled second-order 
ODEs (\ref{e7}) and (\ref{e8}) into a set of four first-order ODEs
\begin{equation}
{dS\over d\theta}=2 \sin\theta (R-H_0-1),
\label{e13}
\end{equation}
\begin{equation}
{dT\over d\theta}={R\sin\theta\over 2\phi}(v_0^2-2 H_0),
\label{e14}
\end{equation}
\begin{equation}
{d\phi\over d\theta}=T \sin\theta 
\label{e15}
\end{equation}
\begin{equation}
{dR\over d\theta}=R\left(-{S\over \sin\theta}-{v_0^2-2H_0\over \phi}
{d\phi\over d\theta} + {v_0^2\cos\theta\over \sin\theta}\right)
\label{e16}
\end{equation}
where the auxiliary functions $S(\theta)$ and $T(\theta)$ are defined
from equations (\ref{e15}) and (\ref{e16}). 

The above equations are supplemented by four boundary conditions, two 
each on the axis and at the equatorial plane. On the axis $\theta=0$,
we demand 
\begin{equation}
\phi\to 0\ \ \ {\rm and}\ \ \ S\equiv \sin\theta\left( -{1\over R}
{d R\over d\theta} -{v_0^2-2H_0\over \phi}{d\phi\over d\theta} +
{v_0^2\cos\theta\over \sin\theta}\right)\to 0.
\label{e17}
\end{equation}
The first condition means that the magnetic flux enclosed by the polar
axis vanishes, and the second that the polar axis contains no line 
mass (see Li \& Shu 1996 for a discussion). At the equatorial plane 
$\theta=\pi/2$, we impose the conditions 
\begin{equation}
{d R\over d\theta}=0\ \ \ {\rm and}\ \ \ {d \phi\over d\theta}=0
\label{e18}
\end{equation}
by symmetry. These translate to the requirements
\begin{equation}
S(\theta=\pi/2)=0\ \ \ {\rm and}\ \ \ T(\theta=\pi/2)=0.
\label{e19}
\end{equation}

Numerically, we integrate the set of first-order ODEs from small $\theta$
to larger values using a Runge-Kutta method. To initiate the integration,
we use the following expansions at small angles that satisfy the boundary 
conditions near the polar axis listed in equation (\ref{e17}): 
\begin{equation}
R=\theta^n (a_0 + a_2 \theta^2 + c_0 \theta^n + ......),
\label{e20}
\end{equation}
\begin{equation}
\phi=\theta^2 \left(b_0-{b_0\over 12}\theta^2 + d_0 \theta^n+......\right),
\label{e21}
\end{equation}
\begin{equation}
S=\theta^2\left[-(1+H_0) + {2a_0\over n+2}\theta^n + ......\right],
\label{e22}
\end{equation}
\begin{equation}
T=2b_0 + {a_0 (v_0^2-2H_0)\over 2nb_0}\theta^n + ......,
\label{e23}
\end{equation}
where $n=4 H_0-v_0^2$ is the exponent of the dominant term in the expansion 
for the density $R(\theta)$. It reduces to the familiar form of $n=4 H_0$ 
for the nonrotating case (Li \& Shu 1996). 

The constants $a_0$ and $b_0$ are free parameters of the expansions. They 
are determined, for any given pair of $H_0$ and $v_0$, by shooting the 
solution towards the equator $\theta=\pi/2$ and matching the equatorial 
boundary conditions listed in equation (\ref{e19}). Other coefficients 
that appear in the expansions are related to these two parameters through
\begin{equation}
a_2={a_0\over 2}\left(1-{H_0\over 3}+{v_0^2\over 12}+{n\over 4}\right),
\label{e24}
\end{equation}
\begin{equation}
d_0={a_0(v_0^2-2H_0)\over 2n b_0(n+2)},
\label{e25}
\end{equation}
\begin{equation}
c_0={a_0 d_0\over b_0}(2H_0-v_0^2).
\label{e26}
\end{equation}

\begin{figure}
    \centering
    \leavevmode
    \psfig{file=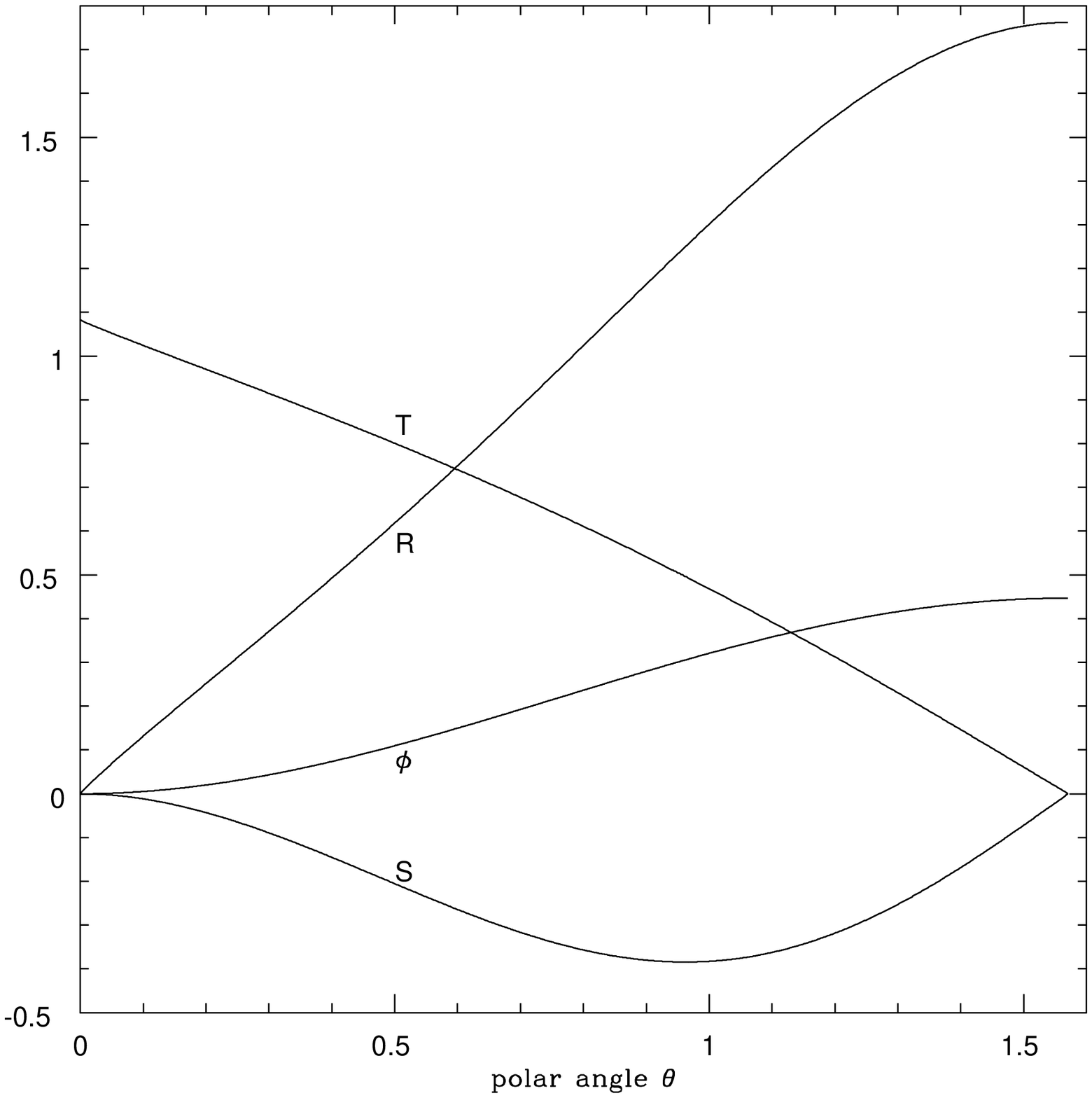,width=0.75\textwidth,angle=0}
\caption{Plotted are the angular distribution functions for density 
$R$, magnetic flux $\phi$, and two auxiliary 
functions $S$ and $T$ against the polar angle $\theta$, for the 
combination of parameters $H_0=0.25$ and $v_0=0.25$. Note that both 
$S$ and $T$ go to zero at the equator $\theta=\pi/2$, which are the
boundary conditions that fix the free parameters $a_0$ and $b_0$ in 
the expansions near the polar axis. } 
\label{fig:f1}
\end{figure}

Equations (\ref{e14}) and (\ref{e15}) imply that the combination
$H_0-v_0^2/2$ is a measure of the angle-averaged support provided by
magnetic fields in comparison with the level provided by thermal pressure.
When this combination vanishes, the magnetic field is zero, and
the pivotal configuration becomes a Hayashi-Toomre toroid.
When this combination is greater than zero, we may introduce a better
measure of magnetization, the
dimensionless mass-to-flux ratio defined in equation (37) of Li \& Shu 
(1996): 
\begin{equation}
\lambda \equiv \int_0^{\pi/2} {R(\theta)\sin(\theta)\over \phi(\theta)}\, 
d\theta.
\label{masstoflux}
\end{equation}
A value of $\lambda = 1$ divides supercritical clouds from subcritical ones.
Since our models have no bounding surface pressures, all cases 
encountered in this paper 
are supercritical, $\lambda > 1$.  For example, the choice $(H_0,v_0)=
(0.25,0.25)$ produces the angular distribution functions displayed in
Fig.~\ref{fig:f1}.  The integral of equation (\ref{masstoflux}) then yields
$\lambda = 4.61$, a case considered by observers to
be magnetically ``weak'' (e.g., Crutcher 1999).  However,
far from satisfying the intuition that magnetic fields are therefore 
ignorable in the 
subsequent collapse, the relatively ``weak" fields in the problem 
actually dominate many
subsequent phenomena of crucial astrophysical interest -- e.g., the 
formation of large
pseudodisks, the transport of angular momentum, and the resulting size of 
centrifugally supported (``true") disks.

\section{Collapse of Magnetized, Rotating Toroids} 
\label{collapse}

Numerically, we follow the inside-out collapse of the rotating toroid 
using Zeus2D (Stone \& Norman 1992), which solves the ideal MHD equations 
in three dimensions with an axis of symmetry. Several modifications 
to the Zeus2D code were needed to follow the toroid collapse; these 
are discussed in Paper I for the nonrotating case. We refer the reader 
to that paper for details. The presence of rotation makes it necessary 
to consider the toroidal component of magnetic field, which is already 
included in Zeus2D. 

\subsection{Collapse of the $H_0=0.25$ and $v_0=0.25$ Toroid}

To illustrate the general
features of the collapse solutions, we first consider as a standard for
comparison the case with $H_0=0.25$ and $v_0=0.25$.  Toroids with 
different combinations of $H_0$ and $v_0$ will be discussed in the next 
subsection. 

\subsubsection{Central Mass Accretion} 

\begin{figure}
    \centering
    \leavevmode
    \psfig{file=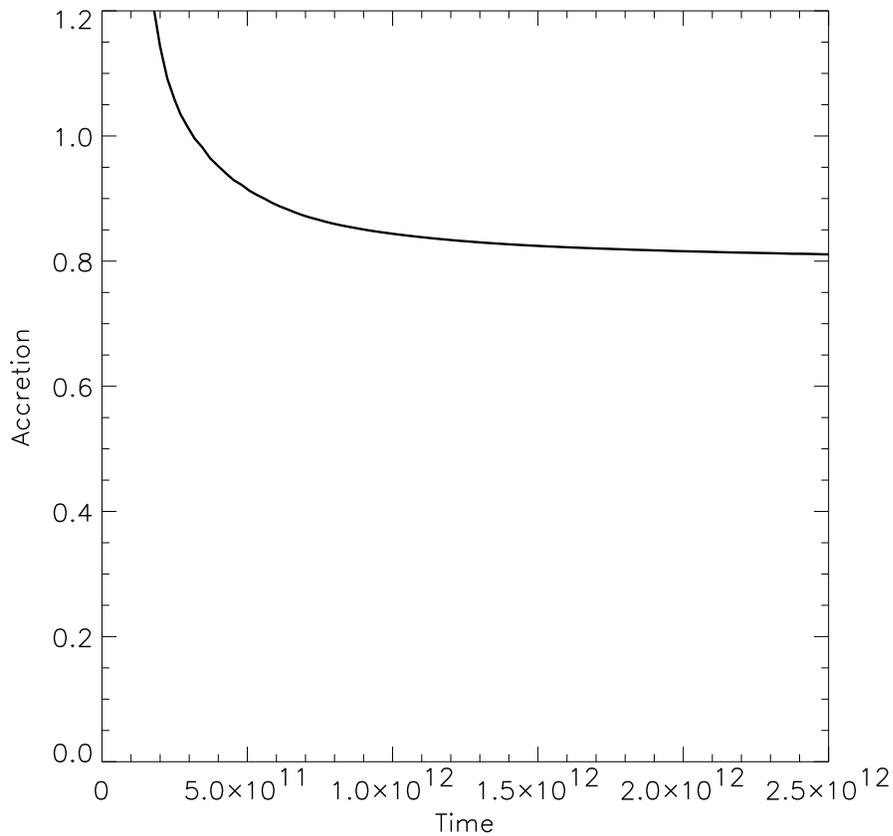,width=0.75\textwidth,angle=0}
\caption{Central accretion rate constant $m_0=G\dot{M}/(1+H_0)a^3$ 
for the fiducial toroid of $H_0=0.25$ and $v_0=0.25$. Here we adopt 
a definition of $\dot{M}$ of the mass in the central cell divided 
by time (which differs from the instantaneous accretion rate shown 
in Fig.~3d of Paper I) to make the curve smooth. The initial peak 
is a numerical artifact from the way that the computation is started.} 
\label{fig:massr}
\end{figure}

We first examine the time evolution of the mass accretion rate into the 
central cell (point mass), which is displayed in Fig.~\ref{fig:massr}.
To facilitate comparisons with observations, we have plotted
the abscissa in physical time units, adopting a fiducial value for the
isothermal sound speed of $a = 0.2$ km s$^{-1}$.  The mass accretion rate
into the central cell is plotted in units of $(1+H_0)a^3/G$, and the 
asymptotic value, $0.80$ is presumably the correct self-similar solution 
(for arbitrary 
values of $a$ and physical times $>$ 0) if one could rid the calculation
of numerical artifices.  Among such artifices is the possession of a 
strong initial peak in the accretion rate, reminiscent of the
peak in the nonrotating case (Fig.~3d of Paper I) induced by
an artificial point mass placed at the center to initiate the inside-out 
collapse. After a short period of adjustment, the central accretion 
rate quickly settles down to the more or less steady value of $\sim 
0.80(1+H_0)a^3/G$. The value is only slightly reduced (by $\sim 18$\%) 
from that of the corresponding nonrotating case (see Paper I). Evidently 
accretion into the central sink cell has not been significantly impeded 
by rotation. 

\subsubsection{Rotating Pseudodisk}

The most prominent feature of the nonrotating collapse solution is the 
formation of a dense, flattened pseudodisk in the equatorial region 
(Paper I), which collapses radially towards the center. Inclusion of 
a modest amount of rotation does not significantly modify this basic 
density structure, as shown in Fig.~\ref{fig:n1r}. In the figure, the 
initial state is plotted, along with the collapse solution at $t=2.5
\times 10^{12}$~seconds
after the initiation of collapse when artifacts
from the starting conditions have essentially disappeared 
(see Fig.~\ref{fig:massr}).
For the convenience of observers, we have again plotted the results
in physical units using the canonical choice $a = 0.2$ km s$^{-1}$ 
for low-mass cloud cores in Taurus.  (Scalings to different values
for $a$ or to different times $t$ occur by the rules given in eq.~[3]
of Paper I; these rules continue to hold even when $u_\varphi$ and 
$B_\varphi$ are nonzero.)  The pseudodisk grows in the same way as
it does in the nonrotating case. In 
fact, for the inner collapse region (say within a reduced radius of 
$x=r/at < 0.5$ or a dimensional radius of $r < 2.5\times 10^{16}$~cm
at a time $t=2.5\times 10^{12}$~s when $a=0.2$~km~s$^{-1}$), 
the density distributions of the rotating and nonrotating cases are 
practically indistinguishable from each other. 

The main difference lies in the rotation of the pseudodisk, which is 
of course nonzero for the rotating case. In panel (c) of 
Fig.~\ref{fig:n1r}, we display contours of constant azimuthal speed.
They show clearly that the pseudodisk, defined somewhat arbitrarily
as the darkest region in the plot, has a significant rotation, with 
the (linear) rotational speed peaking in the middle section of
the disk, in a ``ridge'' (or a ``shell''in three dimensions) that
extends from the disk upwards into the envelope. The rotational
speed has a maximum value $\sim 50\%$ above the sound speed. It falls
off to subsonic values in regions both closer to the polar axis and 
further away. We note that, despite significant rotation, the motions 
inside the pseudodisk remain predominantly poloidal, as in the 
nonrotating case. 

Another feature absent from the nonrotating case is the the twisting 
of magnetic field lines. In panel (d) of Fig.~\ref{fig:n1r}, we plot 
contours of constant field pitch-angle, defined as 
$\theta_{\rm t}=\tan^{-1}(B_\varphi/B_p)$, where $B_\varphi$ and $B_p$ 
are the toroidal and poloidal field strength. Clearly, (differential) 
rotation has distorted the initial, purely poloidal field configuration, 
although not by much. The maximum twist again occurs in a ``ridge'', 
roughly where the rotation is fastest, with a pitch angle of order 
20$^\circ$. In the polar region interior to the ridge, the plasma $\beta
\ll 1$, and the field lines are essentially rigid against wrapping. 
Outside the ridge, the collapsing flow has yet to be spun up enough 
to wrap the field lines appreciably (even though the plasma $\beta$ 
is greater than unity and would allow large twists). 

A more graphical display of the 3D magnetic field is presented in 
Fig.~\ref{fig:Btwist}.
The view obtained at 60$^\circ$ inclination to the rotational and 
magnetic axis
testifies to the relatively small amount of field wrapping that 
occurs in the problem
because magnetic effects dominate those of rotation in the regions of 
interest.
The pole-on view of the same magnetic field lines allows a clearer look 
at the top half of
the twisted split-monopole configuration within the pseudodisk and 
permits a visualization of the magnetic torques ($\propto \varpi B_\varpi
B_\varphi$, where $\varpi=r \sin\theta$ is the cylindrical radius) that 
lead to efficient angular momentum transfer and magnetic braking.
The lack of significant field winding on scales of the outer pseudodisks
and larger may explain why OH-maser polarization studies 
of magnetic-field directions in regions of massive star formation show 
surprising correlation with the large-scale magnetic field of the Milky 
Way Galaxy on a scale of at least 2 kpc (Reid \& Silverstein 1990, 
Fish \& Reid 2003). The implied accompanying loss of angular momentum
in the gas during collapse may already have been detected in high-mass 
star-forming regions by Keto, Ho, \& Reid (1987).

In the absence of angular momentum transport, we would have expected the
formation of a rotationally supported disk with approximate size $0.25 
\, v_0^2 \, at$, or $\sim 5\times 10^{14}$ cm at the time shown in 
Fig.~\ref{fig:n1r}. No such disk has appeared in the actual calculation,
perhaps because of the lack of adequate spatial resolution in the central 
regions, but more likely because of the efficiency of the magnetic 
braking, as we shall discuss in the next subsection.

One might also question whether the assumption of field freezing is likely 
to hold at small scales where densities are high and ionization fractions 
become very low.  We address this practical question in \S~\ref{discussion}; 
here, we merely remark that a substantial layer of the pseudodisk is 
likely to remain suitably well coupled to magnetic fields almost right 
to the surface of the central protostar and makes equivocal any answer 
to whether or not one can expect the formation of true Keplerian disks 
in the face of efficient magnetic braking.

\subsubsection{Outflow, Magnetic Braking and Angular Momentum 
Redistribution}

A somewhat unexpected feature of the collapse of the rotating toroid is 
the presence of an outflow region, part of which is already apparent 
in the upper-left corner of panel (b) of Fig.~\ref{fig:n1r} from the 
velocity vectors. The highly distorted velocity contours (particularly 
that of $v=1.0$) are another manifestation of its presence. In 
Fig.~\ref{fig:velr}, we 
show the outflow more clearly over a larger region, including all
four quadrants. The
outflow has a relatively low speed, comparable to the sound 
speed $a$.  It is thus unrelated to the famous bipolar molecular outflows
of many CO observations.  The same statement applies to the low-speed
outflows obtained in Tomisaka's (1998, 2002) calculations, although 
Tomisaka expresses contrary sentiments.  Nor
are the motions to be confused with the fast-moving,
hypersonic jet/wind that is thought to be driven from near the 
forming star (K\"onigl \& Pudritz 2000; Shu et al. 2000).  Some shaping 
of such winds, if they start out in a wide-angle configuration
(see Shang et al. 2002), might occur,
if they, like their low-speed counterparts in Fig.~\ref{fig:velr}, 
escape preferentially along the low-density, axial direction of
the ambient medium.  In any case, interaction between the 
fast-moving jet/wind of an embedded protostar with the transonic 
outflow/infalling envelope appears unavoidable, but we will postpone 
a treatment of this interaction to a future investigation. 

The slow outflow is driven by magnetic braking. As collapse proceeds, 
matter in the magnetically dominated polar region (where the plasma 
$\beta < 1$) first slides along field lines towards the equatorial 
plane to create the dense, pseudodisk. The pseudodisk is not an 
equilibrium structure; its thermal and magnetic forces are too weak to
balance the gravitational pull of the central object. As the material
in the disk falls inward dynamically, it tends to spin up because of 
angular momentum conservation. However, the field lines threading the 
contracting pseudodisk also pass through the more slowly rotating envelope 
material higher up. The twisting of field lines in the pseudodisk generates 
a torsional Alfv{\'e}n wave, which propagates away from the equatorial
plane. The wave removes angular momentum from the pseudodisk and deposits 
it in a small fraction of the envelope material near the $\beta=1$ line, 
which moves away as a low-speed wind. The outflow is concentrated in the 
``ridge'' where the rotation is fastest and the field lines are most 
twisted (see panels [c] and [d] of Fig.~\ref{fig:n1r}). Closer to the
polar axis, the plasma $\beta\ll 1$ and the wrapping of field lines 
unwinds almost instantaneously; further away, the field lines are not 
twisted enough to reverse the collapsing inflow. We will show below 
that, as the field strength increases, the slow outflow grows wider and 
faster.  

The angular momentum redistribution due to magnetic braking is shown
pictorially in Fig.~\ref{fig:n1rangle}. In the figure, angular 
momentum is depleted from the depressed, ``valley'' in the equatorial
region due to 
collapse. In the absence of magnetic braking, all of the depleted 
angular momentum should be deposited in the pseudodisk, where the 
``mountain'' in the figure is. Magnetic braking lowers the 
height of the ``mountain'' considerably, creating a ``ridge'' that
runs more or less parallel to the polar axis. From the volumes 
beneath the ``mountain'' and the ``ridge'', we estimate that more 
than $90\%$ of the angular momentum that would have accumulated in 
the pseudodisk is removed by magnetic braking in this particular 
case, although the fraction depends on the somewhat subjective 
choice of where the ``ridge'' ends and the ``mountain'' begins. 

For ideal MHD, magnetic braking is very efficient. It removes most
of the angular momentum of the infalling material before it enters 
the central cell. As a result, no centrifugally supported, thin disk 
that we could resolve numerically forms throughout the run of the 
simulation. We believe that even with increased numerical resolution 
a rotationally supported disk will not form in the ideal MHD limit 
under the isothermal assumption. The reason is that in any ideal,
self-similar, calculation that has been carried out correctly, 
the origin must contain a point mass that grows linearly 
with time, and this mass must trap a magnetic flux with a dimensionless 
mass-to-flux given by the the original ratio $\lambda$. The trapped 
flux emanates from the center as a split monopole (Galli \& Shu 1993b, 
Li \& Shu 1996), which causes the infalling matter to become more 
magnetically dominant and slowly rotating as it approaches the 
origin. The reason for magnetic domination is obvious. That for slow 
rotation is less so, and can be shown mathematically as follows. 
Written in cylindrical coordinates,
$(\varpi=r\sin\theta, \phi, z)$, the angular momentum equation for our
problem reads
\begin{equation}
{\partial \over \partial t}(\rho j)+{1\over\varpi}{\partial \over \partial 
\varpi} (\varpi \rho ju_\varpi)+{\partial \over \partial z}(\rho j u_z) =
{1\over 4\pi}\left[ {1\over \varpi}{\partial\over \partial 
\varpi}(\varpi^2B_\varpi B_\varphi)+
{\partial \over \partial z}(\varpi B_\varphi B_z)\right],
\label{angmomeq}
\end{equation}
where $j \equiv \varpi u_\phi \equiv \varpi^2\Omega$ is the fluid 
specific angular momentum.
If we integrate over a cylindrical volume of a small radius $\varpi$ 
and infinite $z$ height,
and if we approximate the fluid distribution to be confined to an 
infinitesimally thin
pseudodisk of surface density $\Sigma$, with $\rho(\varpi, z, t) 
= \Sigma(\varpi , t)\delta (z)$, we obtain
\begin{equation}
{\partial J\over \partial t} +2\pi \varpi \Sigma {ju_\varpi}_{z=0} 
= 2\pi\varpi \int_{-\infty}^{+\infty} {\varpi B_\varpi B_\varphi\over 
4\pi}\, dz,
\label{totangmom}
\end{equation}
where
\begin{equation}
J(t)\equiv \int_0^\varpi \Sigma(\varpi, t)j_{z=0} 2\pi \varpi \, d\varpi.
\end{equation}

Equation (\ref{totangmom}) states that the time rate of change of $J$, the 
total fluid angular momentum contained within a circle of radius $\varpi$
in the plane of the pseudodisk, plus the
radial flux of fluid angular momentum across the circumference $2\pi 
\varpi$ of the same circle,
is equal to the torque exerted by the Maxwell stress acting across a 
cylindrical surface
of midplane circumference $2\pi \varpi$ and infinite height in $z$.  
For $B_\varpi$
satisfying a split monopole geometry, the torque on the right-hand 
side of equation (\ref{totangmom}) can be approximated as a negative 
constant times a representative mean value of $|B_\varphi|$ on the 
important part of this cylindrical surface.

Equation (\ref{induceq}) states, on the other hand, that under the 
assumption of strict
field freezing, azimuthal field $B_\varphi$ is generated at a rate
\begin{equation}
{\partial B_\varphi\over \partial t} = \varpi\left( B_\varpi {\partial 
\Omega\over \partial \varpi}
+B_z{\partial \Omega \over \partial z}\right) .
\end{equation}
If for small $\varpi$, $\Omega$ were to approach anything resembling a 
Keplerian field of differential rotation $\Omega \propto \varpi^{-3/2}$ 
in the presence of a point mass $M=\dot M t$ at the origin, then $B_\varphi$ 
would be generated at a highly divergent rate $\propto \varpi^{-5/2}$.
This behavior would lead to divergent negative torques on the right-hand 
side of equation (\ref{totangmom}) and rob the interior of angular 
momentum that it does not possess. The contradiction implies that 
centrifugally supported disks {\it cannot} form in the presence of
frozen-in magnetic fields of typical interstellar origin if they are 
dragged into the central regions of a pseudodisk configuration.  In 
such circumstances, the angular velocity $\Omega$ is more likely to 
reach a nearly constant value at small $\varpi$, saturating the spinup 
seen at large $\varpi$ in the overall gravitational collapse.  This 
combination of behaviors then explains why the {\it linear} velocity 
of rotation reaches a maximum value at some intermediate value of $\varpi$.

Since exterior gas must ultimately carry away the excess angular momentum 
transferred to them by magnetic torques, one way out of the dilemma posed 
above is to suppose that the magnetic fields on the outside can enforce 
strict isorotation on field lines, so that ${\bf B}\cdot \nabla \Omega = 
0$.  In practice this enforcement is possible only in a near-vacuum 
environment where Alfv\'en waves can propagate almost at infinite speed. 
In other words, we can expect braking to become less efficient once the 
toroidal Alfv{\'e}n wave breaks out of the cloud boundary, a development 
neither allowed by our simulation protocol nor theoretically attractive 
as a general solution to the surprising dilemma of {\it too much} outward 
angular-momentum transport.

As we shall discuss in \S~\ref{discussion}, we believe a better 
alternative is ambipolar 
diffusion and/or magnetic reconnection to decouple eventually
the pseudodisk from its attached magnetic field.  This decoupling would 
diminish the braking 
efficiency and perhaps lead to a growing rotationally 
supported disk at some point during the gravitational collapse.  
The attractiveness of this proposal lies in the unified possibility 
of resolving
two problems of long-standing in star formation without invoking
additional physical processes: the loss of the magnetic flux
associated with trapped interstellar fields, and the appropriate 
outward transport of angular momentum
in disks and pseudodisks.  To wit, an unmagnetized, 
rotating cloud will not collapse directly to a central point-like mass, 
whereas an ideally magnetized, nonrotating cloud can collapse to a 
central point-like mass without a rotationally supported disk.  With 
more realistic physics (see \S~\ref{nonideal}),
a magnetized, rotating cloud may collapse to a protostar while forming
a Keplerian disk that could eventually produce a planetary system.  
Alternatively, if the disk or pseudodisk becomes sufficiently massive
in the process, nonaxisymmetric gravitational instabilities may take 
over to yield binary and/or multiple star systems (see, e.g., the 
speculations of Galli et al. 2001).

\subsection{Effects of Rotational Speed on Collapse Solution}

We now explore, for a fixed value $H_0=0.25$ for the over-density
parameter, a range of values for the dimensionless rotation rate: 
$v_0=0, 0.125, 0.25$ and $0.5$.  The corresponding dimensionless
mass-to-flux ratios $\lambda$ of the sequence are 4.51, 4.61, 4.94, and 7.15.
Thus, the sequence not only represents the effects of increasing rates 
of rotation on the collapse, but
also those of a decreasing magnetic field (both as measured against
the support provided by thermal pressure). 
The self-similar results, all obtained by running the simulations long enough
to eliminate unwanted transients, are shown in Fig.~\ref{fig:allss}. 

To begin the discussion, we note that rotation has only a modest effect 
on the density distribution of the collapse solution. The pseudodisk is
slightly thicker for a faster rotation, partially because the infall 
is retarded more by rotation, and partially because of a more vigorous 
outflow (still very slow by bipolar outflow standards, not to mention
stellar jets), which lifts up some of the disk material (see panel [d]). 
The increasing strength of outflow as $v_0$ increases is most evident 
in the shape of the constant (total) velocity contour of $v=1$, which 
gets more distorted for a larger $v_0$. The distortion comes mostly 
from the toroidal component of the velocity, which can be close to 
the sound speed or larger in the magneto-rotationally driven outflow. 
An interesting feature, related to the strength of the outflow and the
strength of the pre-collapse magnetic field, 
is that the magnetically dominated region enclosed by the $\beta=1$
line shrinks as $v_0$ increases. This effect is seen most clearly
in the fastest rotating, most weakly magnetized $v_0=0.5$ case, where 
the enhanced density in the outflow has 
apparently pushed the $\beta=1$ line close to the polar axis. Note 
also the pronounced distortions of the density contours and field 
lines caused by the outflow in this case. 

Rotation decreases somewhat the rate of mass accretion onto the central 
point mass, a trend also present in the calculations of Tomisaka (2002)
and Krasnopolsky \& K\"onigl (2002). In steady state after the initial 
transient peak, the accretion rate is reduced, respectively, by about 
$14\%$, $18\%$ and $33\%$ for $v_0=0.125, 0.25$ and 
$0.5$, in comparison with the nonrotating case $v_0=0$. The 
reduction may be due to a combination of slower infall and mass removal 
in the outflow.  However, the numerical outward diffusion of magnetic field
exacerbated by artificial viscosity mediating shock waves at
the boundaries of the central cell could also help to lower
effective mass-accretion rates.
In any case, for values of $v_0$ approaching the upper limit 
$\sqrt{2H_0}$ that would make the associated magnetic field in the model 
go to zero, we expect the central mass accretion 
rate to go to zero in any axisymmetric calculation that
also suppresses gravitational torques.  By explicit calculation, we verified
the formal vanishing of $\dot M$ into the central cell for
nonmagnetic calculations in which $v_0$ had
a large enough value (e.g., 0.5)
to allow the numerical resolution of a centrifugally
supported disk. 

\subsection{Effects of Magnetic Field Strength on Collapse Solution} 

For a nonmagnetic Toomre-Hayashi toroid, $H_0=v_0^2/2$. The axisymmetric
collapse of such a toroid 
should not directly produce a point mass at the center in principle. 
In practice, the numerical grid is finite, and the central cell will
experience some mass accretion, particularly when the rotational speed
$v_0$ is small. Sooner or later, the numerical artifacts disappear
from our simulations as the outward expansion of infall
occupies more and more computational cells.  The ``central'' accretion 
diminishes drastically, when most of the infall material lands on a 
spatially resolved,
rotationally supported, disk of approximate size $x_d \approx 0.25 v_0^2$,
in dimensionless units of $\varpi/at$. In an axisymmetric simulation,
the accreted material simply
accumulates in the disk because all processes of angular momentum 
removal or redistribution are suppressed in the absence of any
magnetic and gravitational torques.  Eventually the disks become
sufficiently massive that they are gravitationally unstable to axisymmetric
transformation into rings (Toomre's 1964 instability; see also
Goldreich \& Lynden-Bell 1965), a condition that plagues the long-time 
behavior of some of our computed nonmagnetic results.  In a perfect 
self-similar collapse simulation, some inner part of the disk would 
always be unstable by Toomre's criterion, but the difficulty does not 
manifest itself practically until the calculation has proceeded 
long enough for these regions to become sufficiently spatially resolved 
in a finite-difference scheme.  

The magnetized, rotating collapse is qualitatively different 
due to magnetic braking. Angular momentum removal, particularly in the
magnetically dominated region close to the origin (where $\beta < 1$), 
allows infall material to accrete all the way onto the central cell, 
provided that ideal MHD holds.

To begin our exploration of these effects of magnetic fields,
we fix the rotational speed $v_0$ at the fiducial value of $0.25$, and
consider four different values of $H_0=0.03125$ ($\lambda = \infty$), 0.125
($\lambda = 10.0$), 0.25 ($\lambda = 4.94$), and 0.5 ($\lambda = 2.77$).
The $H_0=0.03125$ case satisfies the nonmagnetic relation $H_0=v_0^2/2$, so
the mass-to-flux ratio is infinite. 
This rotating toroid is initially close to being
spherical, containing only $v_0^2/2 \approx$ 3\% more mass
than a singular isothermal sphere. 
It has a ratio of rotational to gravitational energy of $\sim (v_0/2)^2
=1.56\%$, which is typical of molecular cloud cores (Goodman et al. 
1993). Its collapse is followed with Zeus2D, with the magnetic field 
turned off. The results are shown in panel (a) of Fig.~\ref{fig:allss2}. 
The collapse solution resembles that of the singular isothermal sphere shown 
in Fig.~6a of Paper I, except very close to the origin, where flattening 
of the mass distribution because of rotation becomes more prominent.
The expected 
size of the rotationally supported disk, $x_d \equiv \varpi_d/at \approx 
0.016$
is too small to be resolved by our grid, or the scale of the figure,
in this particular example.  However, by keeping track of the mass and angular
momentum that flowed into the central cell, we verified that angular momentum
conservation holds to high accuracy in our modified version of the 
Zeus2D code.

As $H_0$ increases well beyond $v_0^2/2=0.03125$, most of the extra mass 
over the singular isothermal sphere is supported by magnetic fields rather 
than rotation. This is the case for $H_0=0.125, 0.25$ and $0.5$, whose 
collapse solutions are displayed in panels (b)-(d) of Fig.~\ref{fig:allss2}. 
These panels are to be compared with panels (b)-(d) of Fig.~6 of Paper I, 
where the corresponding nonrotating collapse solutions are plotted. In
all cases, the density distribution shows a prominent pseudodisk, and 
the velocity field contains a slow outflow component. The former is basically
a magnetic structure; it increases in size and becomes more flattened as 
the field strength (or $H_0$) increases, as expected. The latter is due 
to a combination of magnetic fields and rotation. We find that, for a 
given $v_0$, as the field strength increases, the outflow becomes faster 
and wider. It also starts at a larger height above the pseudodisk, most 
likely as a result of a faster propagation speed for the torsional 
Alfv{\'e}n wave in the more strongly magnetized case. 

In each of the three magnetized cases, the slow outflow removes the bulk 
of the angular momentum originally associated with the accumulated mass in 
the central cell. Determining the exact fraction of the angular momentum 
removed is somewhat uncertain. As a rough estimate, we divided the 
average specific angular momentum of the central-cell mass by that in the 
nonmagnetic case.  The ratio typically turns out to be of order $10^{-3}$
or smaller in our best resolved examples, and smaller
than 0.1 in almost all cases.  But in some sense the exercise is pointless,
since in a perfectly resolved calculation, where the central object becomes
a point mass, its angular momentum has to go to zero.  The physically
meaningful number is not how much angular momentum has to be transferred
for matter to reach the center, but what is the mass-accretion rate
achieved in accomplishing this feat, with and without magnetization in the 
problem. Complicating a definitive answer to the latter question is the 
numerical instability described already in Paper I that leads to oscillations
of the magnetic field configuration as it becomes severely pinched and 
relaxes by numerically unpinching. Such oscillations become more severe 
at late times in the simulation (defined by when the innermost parts of
the simulation become well resolved), and interfere thus in a fundamental
way with a determination of the time-steady value of the central $\dot M$.

In the interim, we can examine the effects of magnetic braking on rotation
a little more directly. 
In Fig.~\ref{fig:ang}, we plot the contours of constant rotational speed 
for the nonmagnetic and magnetic cases.  Note that the nonmagnetic 
cloud rotates fastest on small scales, and this rotation is braked 
significantly in all of the three magnetized cases, with the pseudodisk 
in the most strongly magnetized case rotating most slowly, as expected. 
A feature common to all magnetized cases is the strong differential rotation
in and around the pseudodisk, with the rotational speed peaking in the
middle sector where the outflowing ``ridge'' originates. The much smaller 
rotational speed in the region close to the axis (interior to the ridge) 
compared with that in the nonmagnetic case provides clear evidence that
most of angular momentum is indeed stripped before mass enters the central
cell. 

\section{Discussion and Conclusion}
\label{discussion}

\subsection{Magnetic Braking and Keplerian Disk Formation}
\label{nonideal}

Formation of centrifugally supported circumstellar disks is a central 
problem in star and planet formation.  The initial 
distribution of angular momentum in the dense core, which collapses 
to form the star and disk, and the angular momentum redistribution 
during the collapse and disk accretion both affect the final outcome.
We have shown that almost all of 
the core angular momentum can be removed in a low-speed outflow 
driven by magnetic braking during the collapse, provided that (1) 
the core is significantly magnetized and (2) the ideal MHD approximation 
holds. We now examine these two conditions from a practical point of view.

The degree of magnetization of dense cores of molecular clouds prior 
to collapse is somewhat uncertain. In the standard scenario for isolated 
low-mass star formation (Shu et al. 1997; Mouschovias \& Ciolek 1999), 
the cores are envisioned to form out of a magnetically subcritical
background cloud (bounded at its surface by significant thermal
or ram or turbulent pressure), with the evolution driven by ambipolar 
diffusion. The cores remain well magnetized, even during the 
supercritical phase of evolution,
when their dimensionless mass-to-flux ratio $\lambda$
is about 2 (Basu \& Mouschovias 1994; Nakamura 
\& Li 2003). This ratio, corresponding to $H_0 -v_0^2/2 \approx 0.5$ (Li \& 
Shu 1996, this paper), is consistent with the field strength obtained for the
well studied cores L1544 (Crutcher \& Troland 2000) and B1 (Goodman 
et al. 1989) from Zeeman measurements, after likely geometric
corrections (Crutcher et al. 1994; Shu et al. 1999; Ciolek \& Basu
2000). Polarization has been detected in the dust continuum emission
from a number of cores (e.g., Ward-Thompson et al. 2000; Matthews \& 
Wilson 2002), which also 
points to a significant degree of magnetization. In addition, the
cores are known to be elongated in general, with a typical aspect
ratio of 2:1 (Myers et al. 1991); the elongation is too large to 
be induced by rotation, and is consistent with the presence of a
dynamically important magnetic field (Li \& Shu 1996). 
Even if the molecular clouds are magnetically supercritical
to begin with and cores form in them dynamically, it would still be 
difficult to avoid having an $H_0-v_0^2/2$ at least as high as 
0.2 to 0.3, which corresponds to a weak field of $\sim 3~\mu$G 
threading a slab of interstellar medium of $A_{\rm V}=1$. Based on 
these arguments, we believe that the dense cores formed in
present-day star-forming clouds should trap enough magnetic flux for
magnetic braking to operate efficiently during gravitational collapse, 
as long as the field remains well coupled to the bulk of the neutral 
core matter.

Good coupling between the magnetic fields and neutral matter is 
maintained through collisions with ions in relatively low-density 
regions of a dark, molecular cloud core. In such regions, the 
primary source of ionization, and the time scale for ambipolar 
diffusion is typically an order of magnitude longer than the dynamic 
time scale\footnote{In the more turbulent, lower density envelope
surrounding a dense core, the time scale for ambipolar diffusion
can be shortened considerably through field fluctuations (Fatuzzo 
\& Adams 2002). The faster ambipolar diffusion may affect the 
angular momentum evolution during the early phase of core formation.
}, which means that there is little time for magnetic flux 
to be redistributed in a dynamically collapsing flow. Thus, ideal 
MHD is a reasonable first approximation in collapse 
calculations. Nishi et al. (1991; see also Desch 
\& Mouschovias 2001) claim that good coupling is lost above a 
volume density of order $10^{11}$~cm$^{-3}$.  Magnetic decoupling 
occurs for densities higher than this critical value, which reduces
the efficiency of the magnetic braking. The decoupling density 
well exceeds the density in most of the pseudodisk, which is 
typically less than $\sim 10^7$~cm$^{-3}$ (see the density contours 
in Fig.~\ref{fig:n1r} for an example), except in one or two zones
right next to the central cell. Therefore, most of the angular 
momentum of the collapsing core material is already removed by magnetic 
braking before the braking is rendered ineffective by decoupling.

Let us now examine the supposition that the gas volume density will 
eventually exceed
the critical value for decoupling. While true in rotationally supported 
protoplanetary disks with similar physical properties to the so-called 
``minimum solar nebula,'' it is not necessarily the case for the 
dynamically collapsing pseudodisks, whose surface densities are 
much lower and provide less shielding against cosmic ray ionization.  
It is an easy matter to estimate the surface density $\Sigma$ of 
a flattened configuration accreting axisymmertically in quasi-steady 
state $\dot M$ at radial speeds that approach free-fall values $\sqrt{ 
2GM/\varpi}$ onto a central mass $M$:  
\begin{equation}
\Sigma = {\dot M\over 2\pi \varpi \sqrt{2GM/\varpi} } = 0.16 \; {\rm g \, 
cm}^{-2} \left( {{\dot M}\over 10^{-6} \;  M_\odot {\rm yr}^{-1}}\right)
\left( {M\ \varpi \over M_\odot \; {\rm AU}}\right)^{-1/2}.
\label{surfdenff}
\end{equation}
Within a factor of 2 or so, the inner regions of pseudodisks obtained by 
numerical simulation in this paper satisfy the estimate of equation 
(\ref{surfdenff}).

Galactic cosmic rays can keep a layer
with a surface density of about 100 g cm$^{-2}$ sufficiently ionized as to
couple well to magnetic fields (Nishi et al. 1991).  Thus, in the absence
of a magnetized YSO wind or jet that can sweep out cosmic rays before 
they penetrate pseudodisks, Galactic cosmic rays will make a pseudodisk 
with a surface density profile given by equation (\ref{surfdenff}) 
sufficiently well-ionized all the way to the surface of a protostar with 
characteristic radius $\sim 10^{-2}$ AU. Even after a YSO jet or wind 
turns on, if the event is accompanied by stellar magnetic activity and
the emission of X-rays, the observed levels of X-rays in low-mass 
protostars are sufficient to keep a layer with the surface-density 
profile of equation (\ref{surfdenff}) marginally well-coupled
(Glassgold, Najita, \& Igea 1997).

Why then, should centrifugally supported disk ever arise?  Because the 
coupling to magnetic fields in the pseudodisks, while good, is not 
perfect.  Nonideal MHD effects of two types have been described in the 
literature that can substantially modify the discussions considered so 
far in this paper.

Li \& McKee (1996) focused on slippage of magnetic flux in the radial 
direction by the effects of ambipolar diffusion.  They proposed
that the magnetic flux decoupled this way from the mass that enters the
star can drive a hydromagnetic shock against the collapsing 
inflow. Behind the shock, the accretion flow can be slowed down
by a large factor, and become magnetically dominated. It is 
conceivable that magnetic braking would still remove a large fraction
of the remaining angular momentum in the slowly-contracting 
post-shock region (Krasnopolsky \& K\"onigl 2002). 

Galli \& Shu (1993b) focused on the vertical stratification produced
by the highly pinched magnetic field in the pseudodisk as a condition 
whereby magnetic flux can be lost by reconnection.
As we also noted in this paper and paper I, infall in the pseudodisk
tends to produce a split-monopole geometry with reversed poloidal 
fields across the midplane. In an ideal MHD calculation, the reversed 
magnetic fields across the midplane are supported by current sheets.  
In reality, the large electric current density near the midplane will 
dissipate by the presence
of nonzero electrical resistivity.  The dissipation of these currents
corresponds to annihilation of the reversed fields above and below the 
midplane.
The higher magnetic pressures of the surrounding regions will then press
toward the middle, leading to further field annihilation.  Except for the
continued introduction of new interstellar flux by infall into the 
pseudodisk, the entire magnetic flux threading the pseudodisk could 
eventually be lost by this reconnection process.

In quasi-steady state, we therefore expect a ``sandwich-like'' 
structure to develop.
The ``bread'' of the sandwich is the well-coupled,
magnetized, regions where angular momentum can be removed at a sufficient 
pace as to allow the matter to spiral essentially unimpeded at dynamical 
rates toward the central protostar.  This bread is the radially inward 
extension of the original pseudodisk, and it may already have been detected
in the case of IRS L1489 by the observations of Boogert, Hogerheijde,
\& Blake (2002).  The ``meat'' of the sandwich 
is the part which is sufficiently shielded by the bread against external 
ionizing radiation as to become essentially decoupled from any magnetic 
fields.  This decoupling causes eventual spinup of the inwardly spiraling 
material, because of the lack of magnetic braking, and the eventual 
formation of a centrifugally supported disk.  This ``true'' disk
is bounded above and below by the dynamically 
infalling remnant of the pseudodisk.  How massive the meat becomes 
relative to the bread is, at this point, entirely a matter of 
speculation.  It could vary considerably between regions of isolated 
star formation,
such as Taurus-Auriga, and regions of crowded star formation, such 
as Orion, because
of the difference of ionizing radiation (particulates and photons) 
associated
with star-birth activity.

Like many examples in astrophysics, moreover, such as the Sun or
other magnetically active stars, the reconnection process
may occur not quiescently but explosively and
intermittently in flares.  It is interesting to speculate whether 
such intermittent behavior has anything to do with FU Orionis outbursts.
Other effects may enter to further complicate the situation.  The 
flow could become turbulent because of Kelvin-Helmholtz instabilities 
that arise 
from the velocity shear between the bread and meat parts of the
sandwich, or because of interchange instabilities in 3D that manifest
themselves in the pinched magnetic configuration of the split-monopole.  
Under laminar conditions, resistive effects become important
in the same density regime where ambipolar diffusion becomes competitive
with the dynamics (Nishi et al. 1991), so a combination
of the effects discussed by Galli \& Shu (1993b) and Li \& McKee (1996)
may be operative.  All these effects operating together in lightly 
ionized media, under turbulent conditions, may considerably increase the
size scales over which nonideal effects must be considered (Zweibel 
\& Brandenburg 1997a, b). 
Challenging nonideal 3D simulations with adaptive mesh refinement 
may be needed to quantify such possibilities. 

\subsection{Comparison with Previous Work} 

Magnetic braking in star formation has been studied previously 
by many authors, in connection with the long-standing ``angular 
momentum problem''. Most of the early studies were concerned 
with the relatively low-density, initial stage of star formation 
using semi-analytic treatments (e.g., Mestel \& Paris 1984; 
Nakano 1989; see Mouschovias 1990 for a review). Recent numerical 
calculations have extended the treatment to higher densities 
associated with the runaway core contraction either prior to star 
formation (Basu \& Mouschovias 1994) or slightly beyond the 
formation of a central object (Tomisaka 2002). These calculations 
showed that, during the short runaway phase when the central density
increases by several orders of magnitude, the angular momentum 
is nearly conserved, because the density increase involves little 
positional change of fluid elements in both radial and azimuthal 
directions (and thus little pinching and twisting of field lines). 
Our work deals 
exclusively with magnetic braking in the infall or accretion phase, 
{\it after} the formation of a compact central object. One may naively 
think that, once a molecular-cloud core starts to collapse dynamically, there 
should be little time for magnetic braking to operate. And yet we 
find that the bulk of the angular momentum, if not the total, can 
be removed during the dynamical collapse, even with a magnetic field 
of relatively moderate strength. Why is there this difference between 
previous expectations and current calculations?

There are two key ingredients behind the efficient braking 
during the accretion phase: (1) the domination of magnetic
energy over the rotational energy, and (2) the distinct geometry of 
the magnetic field. The foot points of the magnetic field lines 
associated with the matter already landed on the central object 
are pinned near the origin by accretion, creating a split monopole for
the magnetic geometry (see Fig.~\ref{fig:n1r}). The split-monopole 
field lines push on the field lines outside (see also Li \& Shu 1997),
bending them into an 
hour-glass shape that is conducive to efficient magnetic braking.
The enhancement comes from both the greater field strengths due 
to pinching and the radial fan-out (as opposed to vertical 
collimation) which lengthens the level arm for magnetic braking. 
These two effects are not taken into account in previous simple
estimates. They are analogous to magneto-centrifugally driven 
winds (K\"onigl \& Pudritz 2000; Shu et al. 2000). Indeed, the 
slowly moving outflow we find in the collapse solution is 
fundamentally similar to these winds, except for scale and for
the ultimate origin of the field (interstellar, disk dynamo, or
stellar dynamo). 

Magneto-rotationally driven outflows were also found in the collapse 
calculations of Tomisaka (1998, 2002). These calculations start 
from infinitely long cylinders with magnetic field lines parallel 
to the axis. The magnetized cylinder is induced to break up into 
dense cores by a sinusoidal density perturbation along its length. 
The subsequent core evolution prior to the formation of a central 
object is more dynamic than that driven by ambipolar diffusion. It 
produces a pivotal state at time $t=0$ (when a {\it hydrostatic} core 
starts to form in the calculation) that differs substantially from 
the singular toroids that we adopted based on ambipolar diffusion 
calculations. Nevertheless, soon after the appearance of the hydrostatic 
core, an outflow develops, which Tomisaka attributed to core formation. 
A more precise cause for the outflow, we believe, is the trapping of 
a finite amount of magnetic flux (rather than mass) in a small (core) 
region, which gives rise to the central magnetic monopole responsible 
for the pinched magnetic geometry that is conducive to outflow 
driving. For numerical reasons, Tomisaka's calculations were stopped 
a short time into the accretion phase, typically $4\times 10^{-3}$ 
times the free-fall time at the center of the initial cylinder beyond 
the moment of hydrostatic core formation (t=0). During this brief 
period of time, the outflow appears to have removed most of the  
angular momentum of the material close to the origin (Tomisaka 2000). 
In particular, no rotationally supported disk is able to form,  
consistent with our findings which, because of their self-similar
nature, apply to all times $t>0$.    

Formation of rotationally supported disks during the collapse of 
dense cores formed in strongly magnetized molecular clouds was 
recently examined by Krasnopolsky \& K\"onigl (2002) under thin-disk 
approximation. The material in the disk is connected through field 
lines to an external medium, which brakes the disk rotation with a 
parameterized efficiency. They find that, in the ideal MHD limit, a 
rotationally supported disk can form, provided that the braking 
efficiency is low enough (see their Fig.~3). If the braking is more 
efficient, the disk angular momentum can be radiated away completely, 
as shown in their Fig.~6. Our calculations and those of Tomisaka 
(1998, 
2002), which treat the dynamics of the material above and below the 
disk self-consistently, suggest that the braking efficiency is high 
enough to prevent the rotationally supported disk from forming in
a strongly magnetized core, as long as ideal MHD holds. In the 
presence of ambipolar diffusion, Krasnopolsky \& K\"onigl (2002) 
find that the bulk neutral material 
can retain a small rotation even in the strong braking case, because 
the neutral component must rotate relative to the charged component
in order to feel the magnetic braking torque (see their Fig.~10).  
Whether the residual angular momentum is large enough in realistic
situations to form a rotationally supported disk well outside the 
forming star remains to be seen. 

To conclude this subsection, we note an interesting feature of 
regarding numerical errors, which are always present, of course, 
in any simulation. Simulations of self-similar configurations 
such as ours confer the advantage that although one does not know 
(in analytically intractable situations) what the exact details 
of the solution should be, one does know from dimensional 
considerations how each quantity should scale.  This knowledge 
provides a powerful check on the present simulations, and 
therefore we have a good handle on the numerical reliability 
of the results quoted in this paper.

\subsection{Observational Implications: The Case of IRAM 04191}

Rotation introduces a new feature into our collapse solution - the
slow outflow (see also 
Tomisaka 1998, 2002). Despite its  crucial role in angular momentum 
redistribution, this outflow is unlikely to be observed soon: it 
is too easily masked 
by the faster-moving bipolar molecular outflows driven by a jet/wind 
from near the central object. Another prominent feature of the collapse 
solution -- the flattened, high-density pseudodisk in the equatorial 
region -- should be less affected by the bipolar outflow. Indeed, there
are many examples of dense structures surrounding deeply embedded YSOs 
that are elongated perpendicular to the outflows. They can be plausibly 
interpreted as the pseudodisks formed 
during the collapse of magnetized toroids or, on a larger scale, the
more static toroids themselves, 
although detailed kinematic studies and direct measurements of the
strength and morphology of magnetic fields are needed to strengthen
the case. 
The best studied circumstellar
structure of this type is perhaps the one around the source IRAM 04191 
(Belloche et al. 2002), which we discuss in some depth below.

\begin{figure}
    \centering
    \leavevmode
        \epsfxsize=.48\textwidth \epsfbox{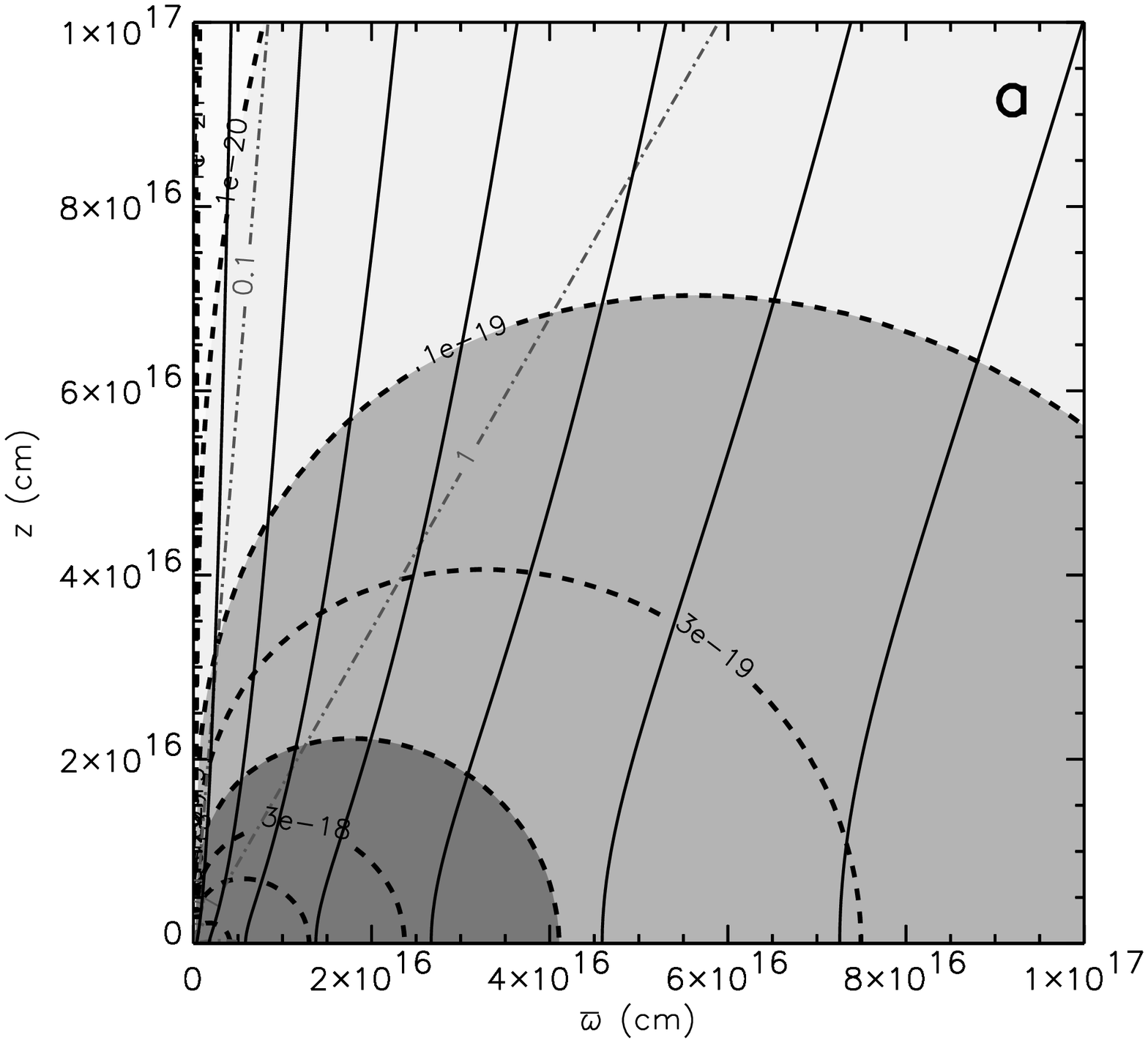} 
        \epsfxsize=.48\textwidth \epsfbox{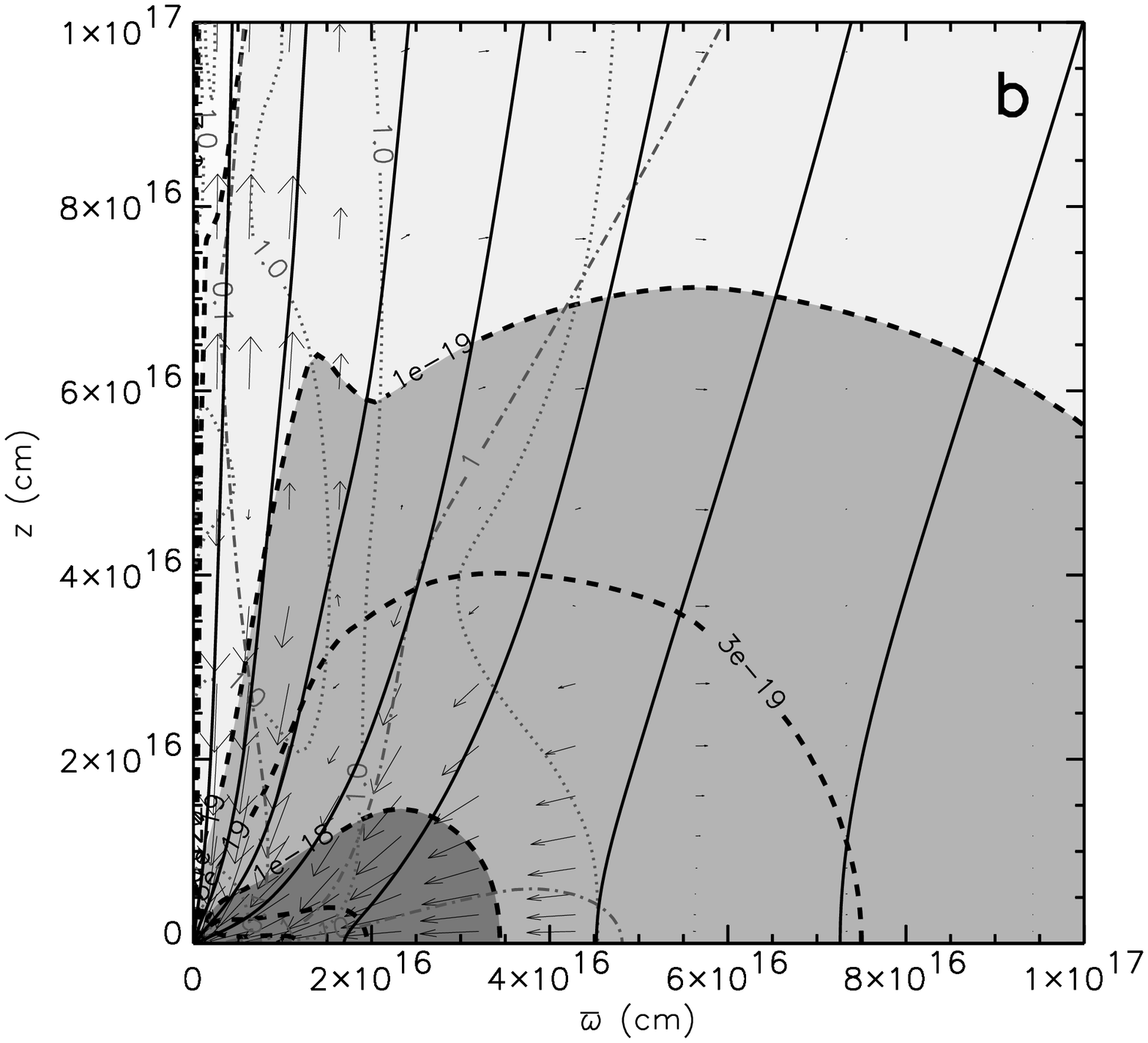} 
        \epsfxsize=.48\textwidth \epsfbox{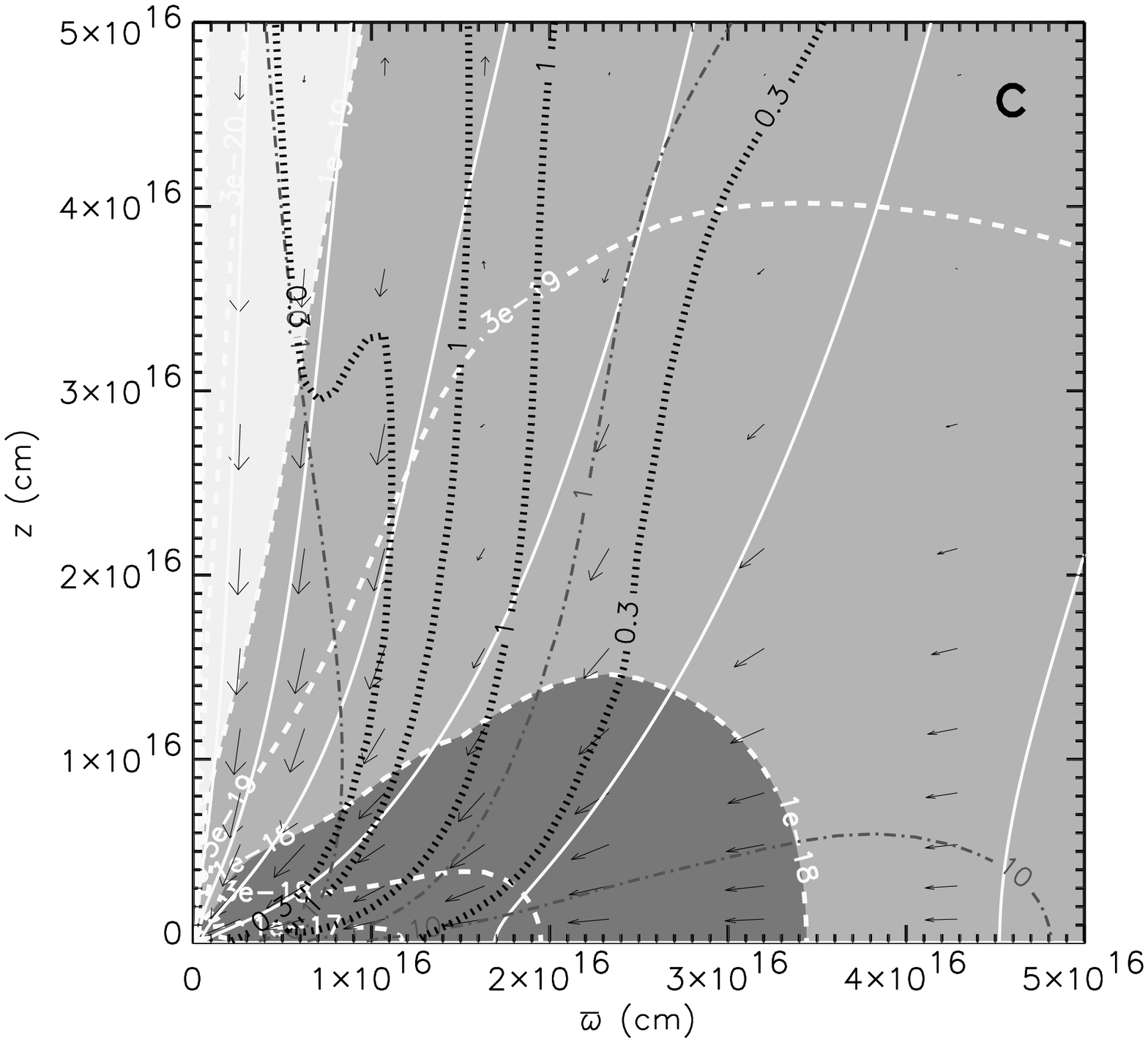} 
        \epsfxsize=.48\textwidth \epsfbox{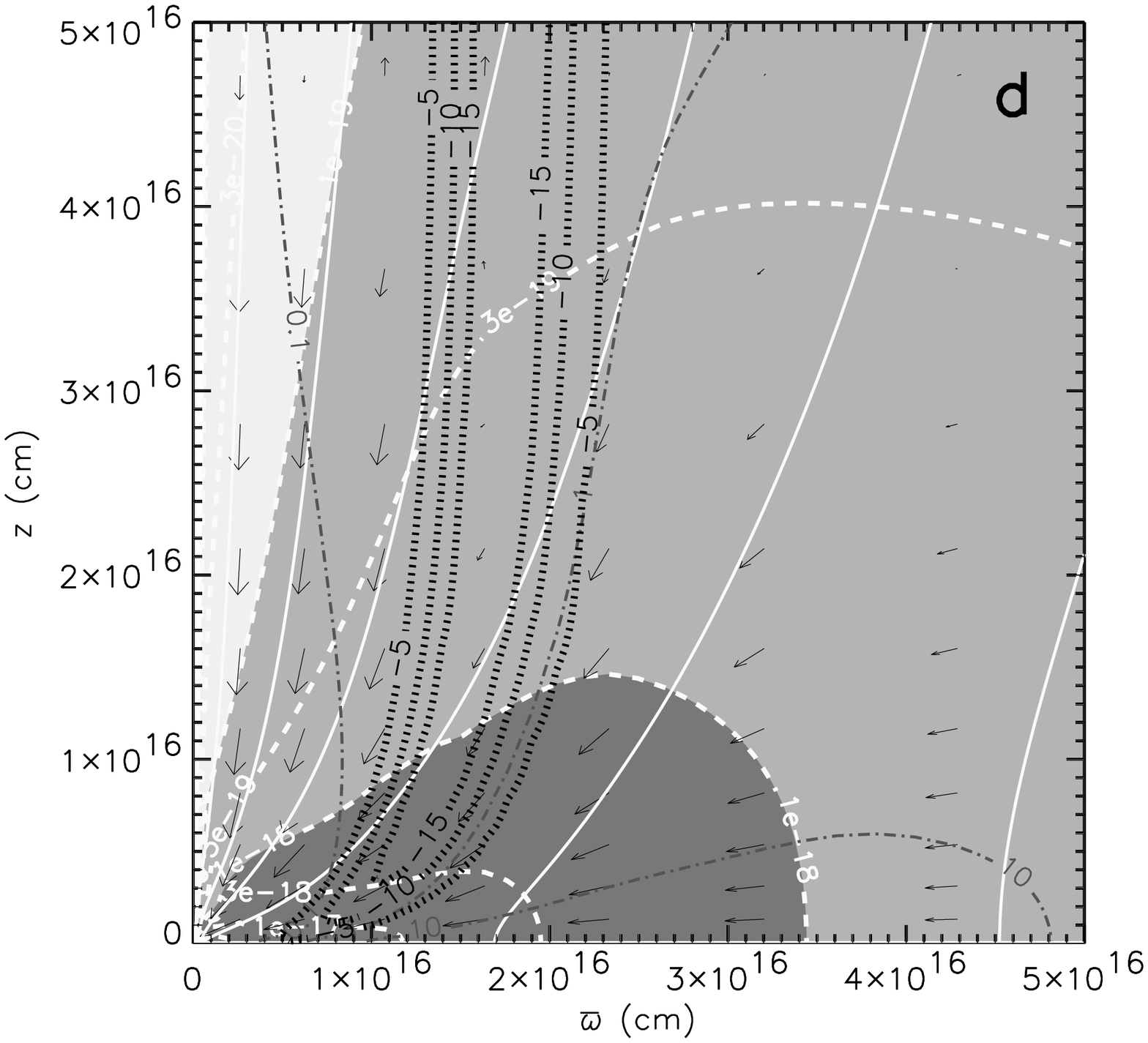}
\caption{(a) The initial state of the fiducial $H_0=0.25$ and $v_0=0.25$ toroid
at $t=0$. (b) The collapse solution at time $t=2.5\times10^{12}$~s. 
The contours of constant density (in units of g~cm$^{-3}$) are plotted as 
dashed lines, with the shades highlighting the high density regions. The 
magnetic field lines are plotted as solid lines, with contours of constant 
$\beta$ (dash-dotted) superposed. The velocity is shown by unit vectors, 
with its magnitude in units of the sound speed $a$ given by the dotted 
lines. Panels (c) and (d) are closeup views of the inner region of panel 
(b), showing the contours of constant rotational speed and pitch angle of 
magnetic field, respectively, in heavy dotted lines.}
\label{fig:n1r}
\end{figure}

IRAM 04191 is a low-luminosity ($L_{\rm bol}\sim 0.15 L_\odot$) protostar
in the nearby Taurus molecular cloud, surrounded by a massive envelope 
($M_{\rm env}\sim 1.5 M_\odot$; Andr{\'e, Motte, \& Bacmann 1999). This 
Class 0 source drives a well-developed molecular outflow, as all Class 0 
sources do. Perpendicular to the outflow lies a flattened structure that 
is mapped in both dust continuum and several molecular lines (Belloche 
et al. 2002). The line study revealed that the flattened structure is 
infalling at a velocity between half and one isothermal sound speed, with 
a substantial amount of rotation. The (linear) rotational speed
increases from $\sim 0.2\ a$ at a radius of $\sim 11,000$~AU
to $\sim a$ at $\sim 3,500$~AU, where $a\approx 0.2$~km~s$^{-1}$ is the 
isothermal sound speed at 10~K (see Fig.~12c of Belloche et al. 2002). 
We identify the increase 
in rotational speed with the spinup in the outer part of the pseudodisk 
as a result of radial contraction under the approximate conservation of
specific angular momentum, before significant magnetic braking sets in.
In our model, the rotational 
speed peaks at some radius, and then decreases inward in a region of 
strongly pinched magnetic field that efficiently removes the angular 
momentum of 
the accreting flow, as shown in Fig.~\ref{fig:n1r}c. 
If the flattened structure of IRAM 04191
is indeed a pseudodisk, then its rotational speed should decrease inward 
inside $\sim$ 3,500 AU according to our model.
This is consistent with, but not required by,
the available spectral data, particularly the ``S''-shaped position-velocity
diagrams of several molecular species. Higher resolution observations are 
needed to test the model.   The test may be hampered by the freeze-out 
of molecules onto dust grains in the high-density region near the 
protostar, as evidenced by the ``hole'' in the NH$_3$ distribution around
this source (Wootten et al. 2003). 

Interpreting the flattened structure of IRAM 04191 strictly in terms of 
the models of this paper is not without difficulties. In the fiducial 
case where
$v_0=0.25$ (comparable to the $0.2 \ a$ seen at 11,000 AU in IRAM 04191), 
the rotational speed achieves a peak near the observed value $\sim a$
at a reduced radius of $x_{\rm p}\sim 0.25$. 
If this corresponds to the radius of $3,500$~AU for IRAM 04191, then the 
collapse would have been initiated $\sim 3.3\times 10^5$~yrs ago, and 
the central star would have a mass of $\sim 0.8 M_\odot$.
But the dynamical time for the molecular outflow of IRAM 04191 is 
only $\sim 8\times 10^3$~yrs, much shorter than the above estimate. The 
low luminosity 0.15 $L_\odot$ of the central source also suggests a 
fairly small mass for the central
star, accreting gas at not much higher than the
standard rates for Taurus (see, e.g., Adams, Lada, \& Shu 1987). 

A flawed way to reconcile theory and observation,
within the context of quasi-steady flow, is to postulate
a significant temporal offset between the pivotal instant $t=0$
and the appearance of a powerful outflow in the system.  Such an 
offset requires,
for example, some way to store infalling material (from a pseudodisk)
in a rotationally supported disk without producing either much accretion
luminosity or an outflow for a duration approaching $10^5$ yr.  This task
is not easy to do without inducing fierce nonaxisymmetric gravitational
instabilities that would undermine the assumption of little mass 
accretion into
the central regions.  Moreover, the observational limit claimed for the 
mass or size of any disk in IRAM 04191 is less than $\sim 10^{-3}$ 
$M_\odot$ or 10 AU.  Efficient magnetic braking has evidently prevented 
a more massive/larger disk from forming close to the star, and such 
braking works against the idea of appreciable mass storage in a disk.

A better solution might be to suppose that, in addition to an offset between
the pivotal instant and the onset of a wind,
accretion onto the central source
is time-variable, with FU Orionis outbursts punctuating quiescent periods 
of relative calm (Hartmann, Kenyon \& Hartigan 1993).  In such a picture,
IRAM 04191 has just recently experienced such an outburst (accounting
for the associated powerful molecular outflow) and has now faded considerably
in its accretion luminosity (accounting for the low luminosity currently).  
Oscillatory inflow into the central regions in the presence of steady infall 
from the pseudodisk is a common bane of our simulations (see the discussion 
in Paper I). It remains to be seen whether the numerical artifact of field 
pinching and unpinching observed in the simulations might not have their 
physical analogues in more realistic calculations performed with nonideal 
MHD.

Another plausible option invokes a pre-existing contracting and rotating 
toroid to explain the observations of the regions at a few thousand AU 
and beyond. Subsonic cloud contraction prior to the formation of a 
formal density singularity for $t < 0$ is consistent with the observed 
motions in both IRAM 04191 and the well-studied starless core L1544 
(Tafalla et al. 1998). 
As discussed already in Paper I, subsonic contraction of the requisite 
magnitude might be initiated by the dissipation of turbulent support in 
dense molecular-cloud cores (Myers 1999) or resulted from ambipolar 
diffusion (e.g., Ciolek \& Basu 2000). We have performed simulations 
of the fiducial (rotating and magnetized) case with an initial uniform
radial inflow $u_r$ equal to half the sound speed $a$.  As we found in 
Paper I for
the example without rotation, the results look like the configuration started
without initial radial motion, except that the
density profiles next to the axis are partially filled in by the additional
inflow and the central accretion rate is increased by about a factor of 2.

\subsection{Conclusion}
\label{conclusion}

To summarize, we have studied numerically the inside-out collapse of 
rotating, magnetized 
singular isothermal toroids. We find that a pseudodisk forms, just
as in the nonrotating case, through which most of the core material 
is accreted. The pinched magnetic field in the pseudodisk removes 
most of the angular momentum of the accreted matter, and deposits 
it in a low-speed wind. The efficient magnetic braking during the 
protostellar accretion phase has important implications for the 
formation of rotationally supported disks that need to be pursued 
with nonideal MHD calculations.

The complementary side to this phenomenon is the discovery of an extremely
simple and efficient mechanism of angular momentum transport that allows 
direct accretion
of extended rotating material onto relatively compact objects without the
necessary concomitant formation of a standard viscous accretion disk.
Whether this mechanism has anything to do, for example, with the mini-spiral
that seems to be feeding ionized gas, with low radiative efficiency,
into the massive black hole at the center of our 
own Galaxy (Lo \& Claussen 1983) remains to be seen.
We have speculated in this paper that 
the process, with the inclusion of nonideal MHD effects, opens new vistas
by which the two classical conundrums of star formation 
discussed long ago by Mestel and Spitzer -- the angular-momentum problem and
the magnetic-flux problem - might be resolved, perhaps simultaneously, 
in the disks and pseudodisks surrounding newly formed stars.}

\acknowledgments{We thank A. K\"onigl and R. Krasnopolsky for useful
correspondence. Support for this work was provided in Taiwan in part 
by grants from
Academia Sinica and the National Science Council, and in the United 
States by grants from the
National Science Foundation and NASA.}

\clearpage

\begin{figure}
\caption{Magnetic field lines for the fiducial case plotted at viewing angles
of 60$^\circ$ and $0^\circ$ (i.e., pole-on) inclination with respect to
the magnetic pole.  Innermost field lines shown in red, blue, and yellow 
are tied
at the origin and pass through the pseudodisk as a split monopole 
configuration.
Filed lines that thread through the midplane and have not yet been 
swept into the central mass point are shown in black.}
\label{fig:Btwist}
\end{figure}

\begin{figure}
    \centering
    \leavevmode
   \psfig{file=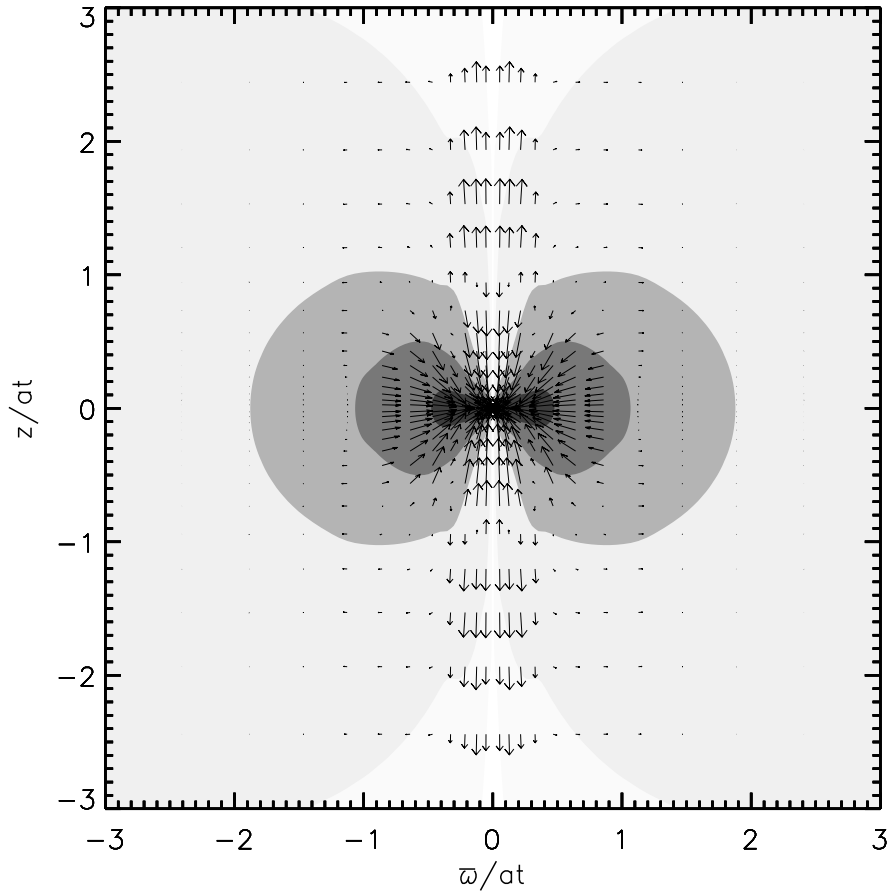,width=0.80\textwidth,angle=0}
\caption{Velocity unit vectors (shown for every fifth cell) for the 
fiducial $H_0=0.25$, rotating toroid.  Unit vectors are three 
dimensional; i.e., the rotational component pointing out of or into the
plane of the page is not shown; thus, the displaced ``unit vectors" 
have different lengths when they are projected into the meridional 
plane. The light, medium, and dark contours extending to approximately
dimensionless cylindrical radius $\varpi \approx 0.5, 1,$ and $2$, 
respectively, correspond to reduced density $\alpha = 10, 1,$ and 
$0.1$, respectively.} 
\label{fig:velr}
\end{figure}

\begin{figure}
    \centering
    \leavevmode
\caption{Angular momentum redistribution due to magnetic braking. Plotted
is the difference in angular momentum between the collapsed and initial
configurations in similarity coordinates, $4\pi G t 
\rho u_\varphi \varpi/a^2$, in the $\varpi/at-z/at$ plane, weighted by a 
factor 
of $\varpi/at$ due to difference in ring volume at different radii. The 
quantities
$4\pi G t \rho u_\varphi \varpi/a^2$, $\varpi/at$, and $z/at$ are shown on 
linear scales.  The volume under the surface in a given region coordinate 
is proportional to the net change in angular momentum of that region from 
the initial pre-collapse state at $t=0$. }
\label{fig:n1rangle}
\end{figure}

\begin{figure}
    \centering
    \leavevmode
        \epsfxsize=.48\textwidth \epsfbox{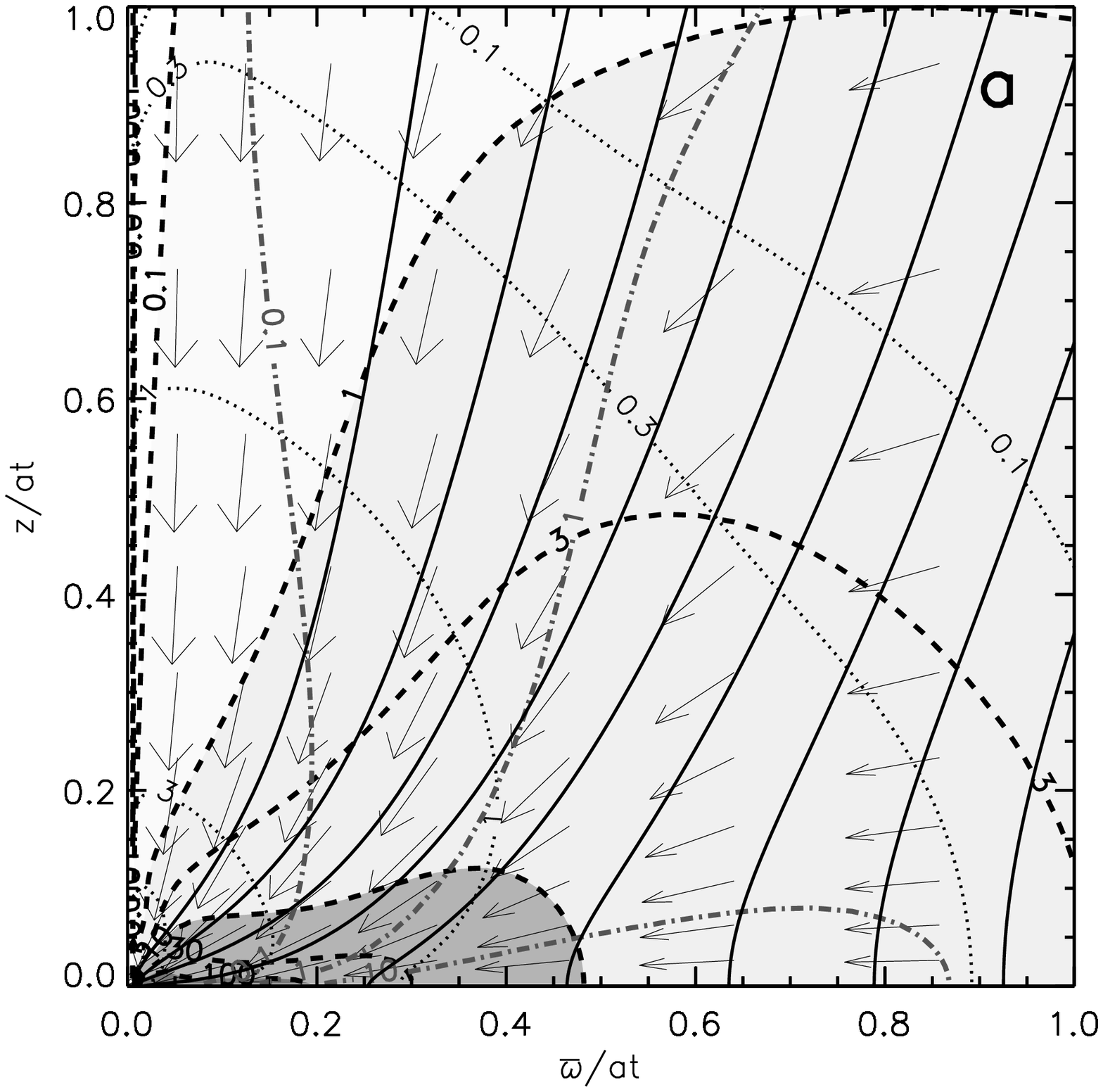} 
        \epsfxsize=.48\textwidth \epsfbox{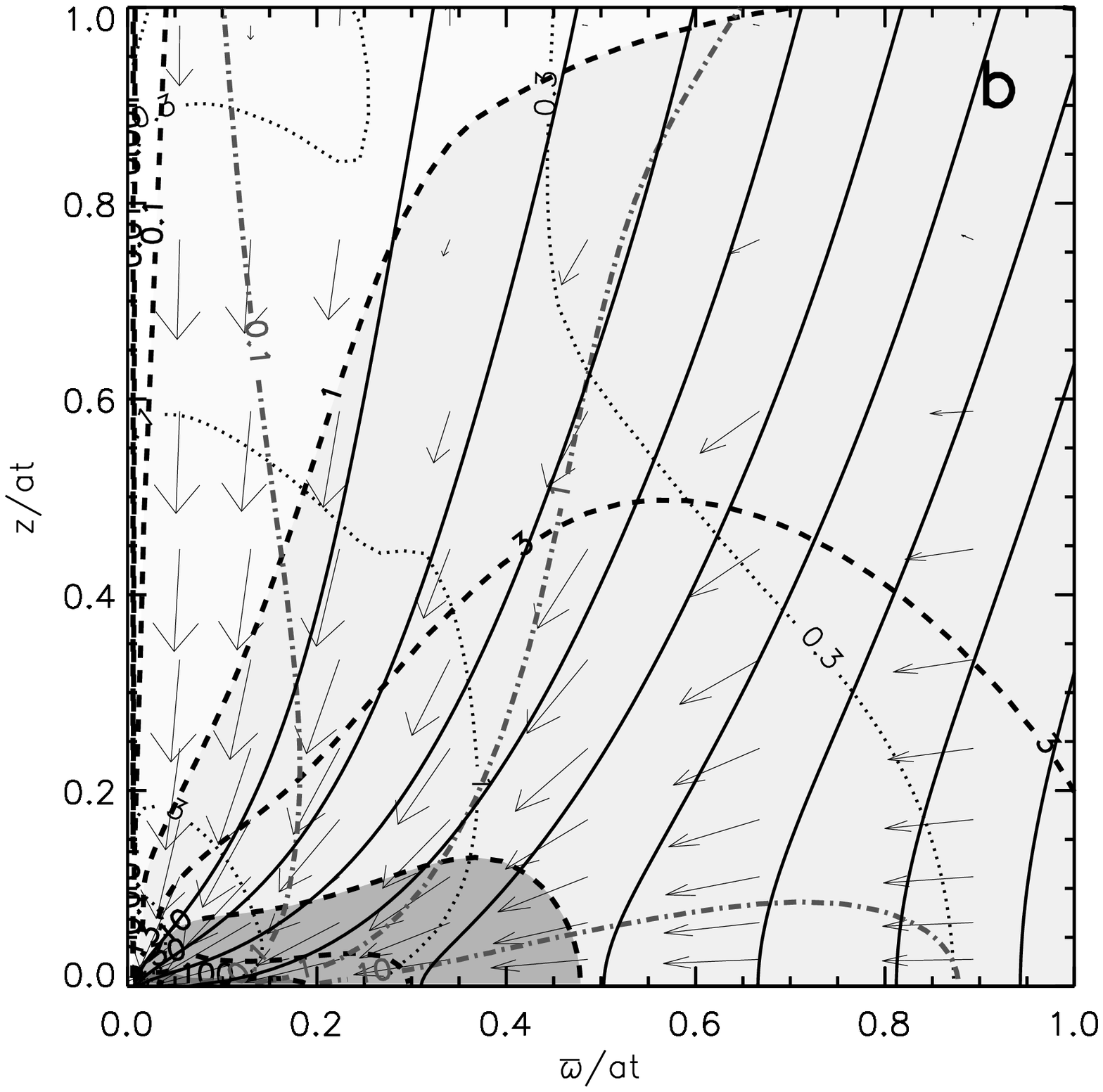} 
        \epsfxsize=.48\textwidth \epsfbox{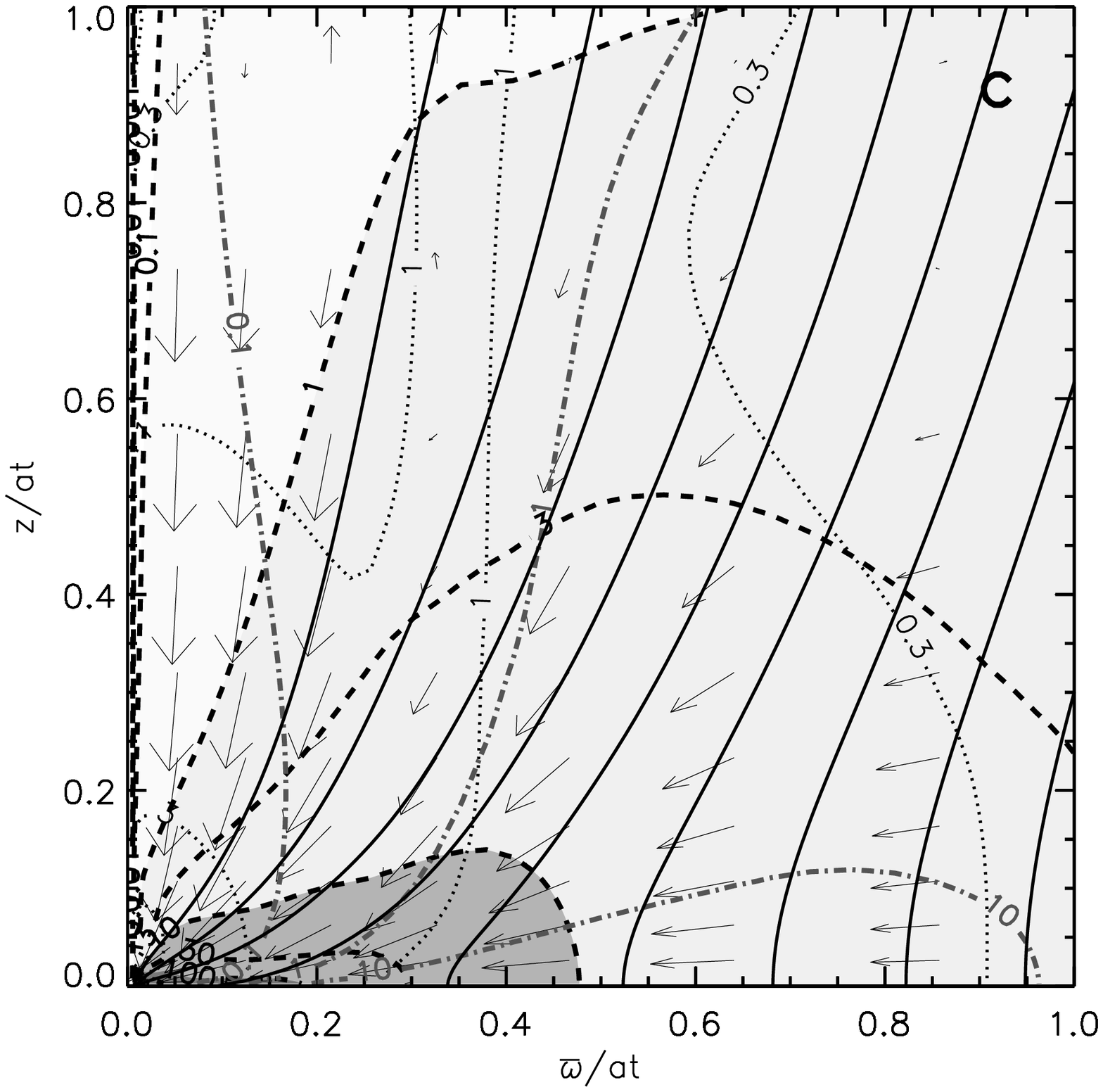} 
        \epsfxsize=.48\textwidth \epsfbox{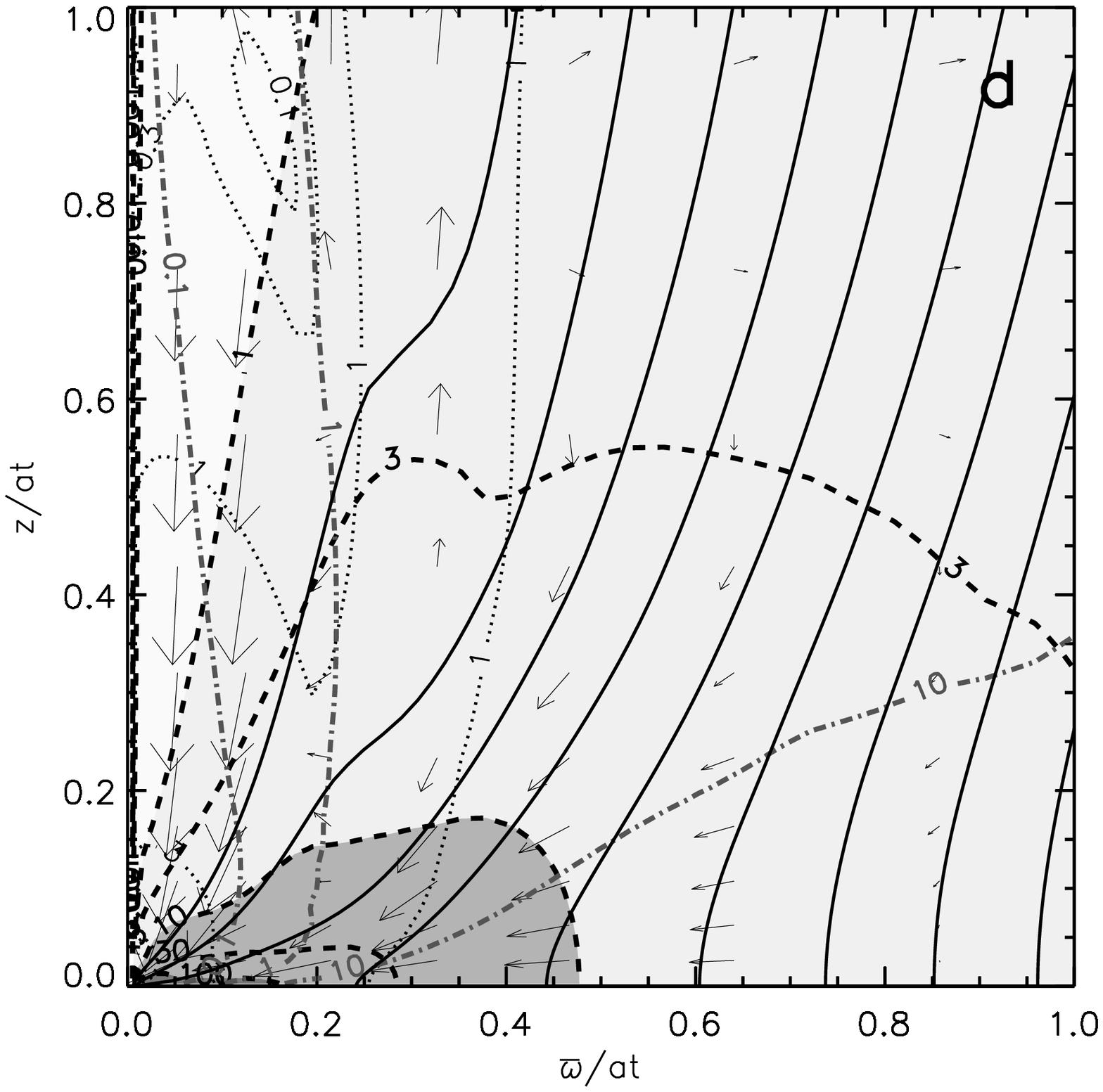}
\caption{Plots of self-similar density, $4\pi Gt^2\rho$, velocity field, 
${\bf u}/a$ and magnetic 
field, ${\bf B}G^{1/2}a/t$, lines for $H_0=0.25$ and $v_0=0, 0.125, 0.25,$ 
and $0.5$.  The isodensity 
contours are plotted as dashed lines, with the shades highlighting the
high density regions. The magnetic field lines are plotted as solid lines,  
with contours of constant $\beta$ (dash-dotted) superposed. The
velocity in every fifth cell is shown by unit vectors, with its magnitude 
given by the
dotted contours. 
Field lines are not the same across all figures; examine the $\beta$ 
contours for field strength.  Notice that a centrifugally supported disk
of dimensionless size $x_d \approx 0.25 v_0^2 \approx 0.06$ when $v_0 = 0.5$
in a nonmagnetic calculation has not appeared in the magnetized calculation
of panel d.}
\label{fig:allss}
\end{figure}

\begin{figure}
    \centering
    \leavevmode
        \epsfxsize=.48\textwidth \epsfbox{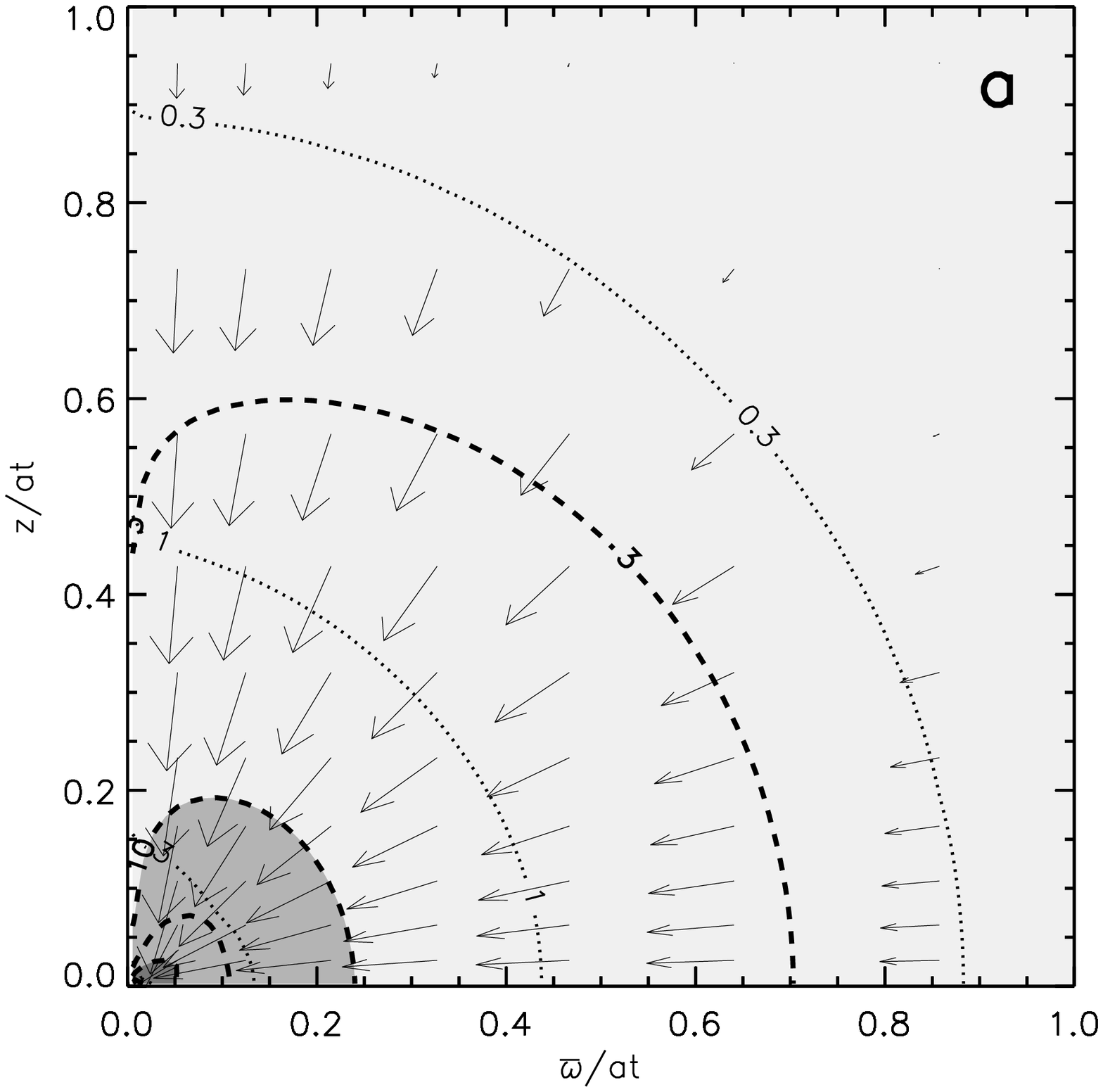} 
        \epsfxsize=.48\textwidth \epsfbox{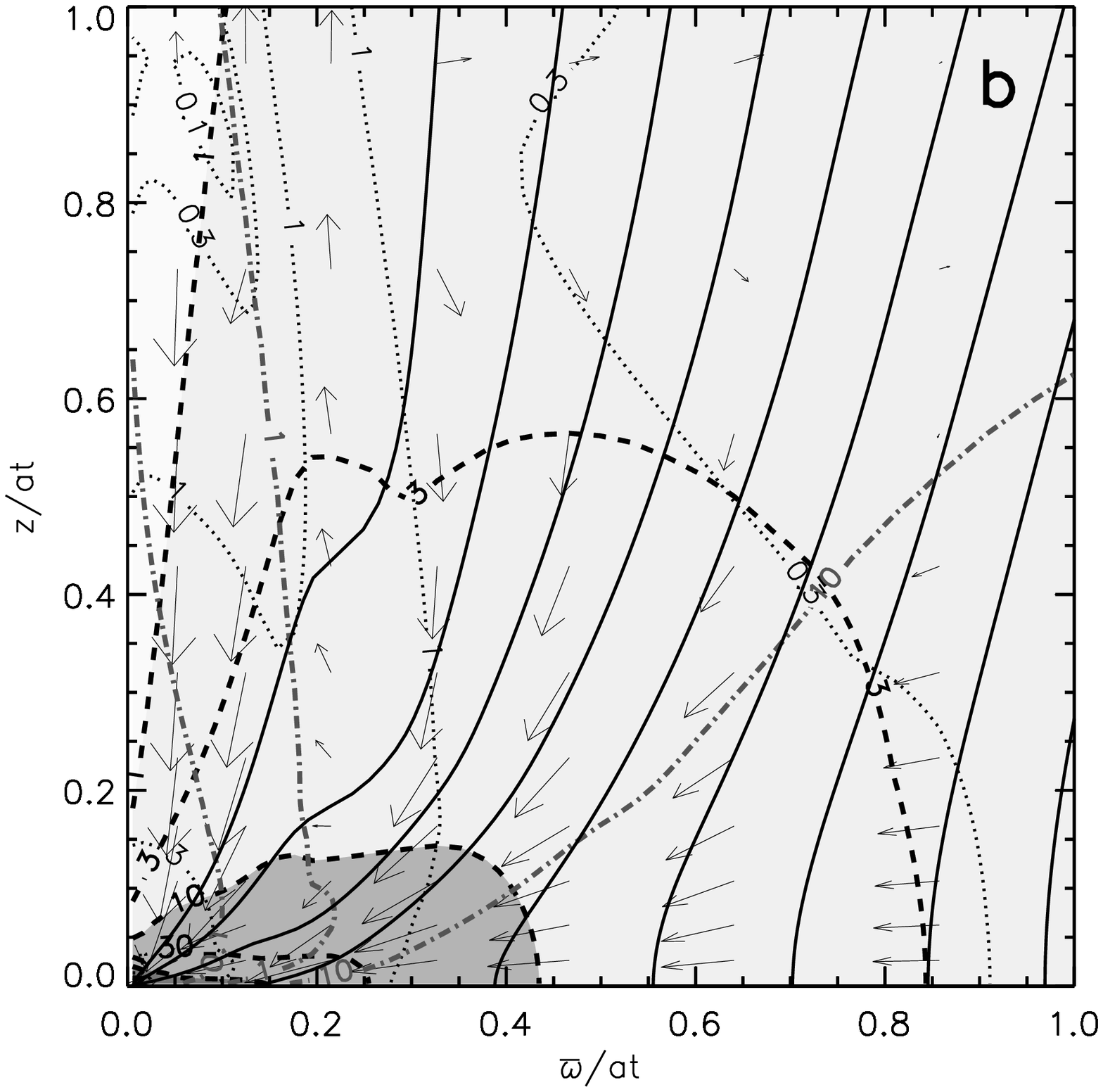} 
        \epsfxsize=.48\textwidth \epsfbox{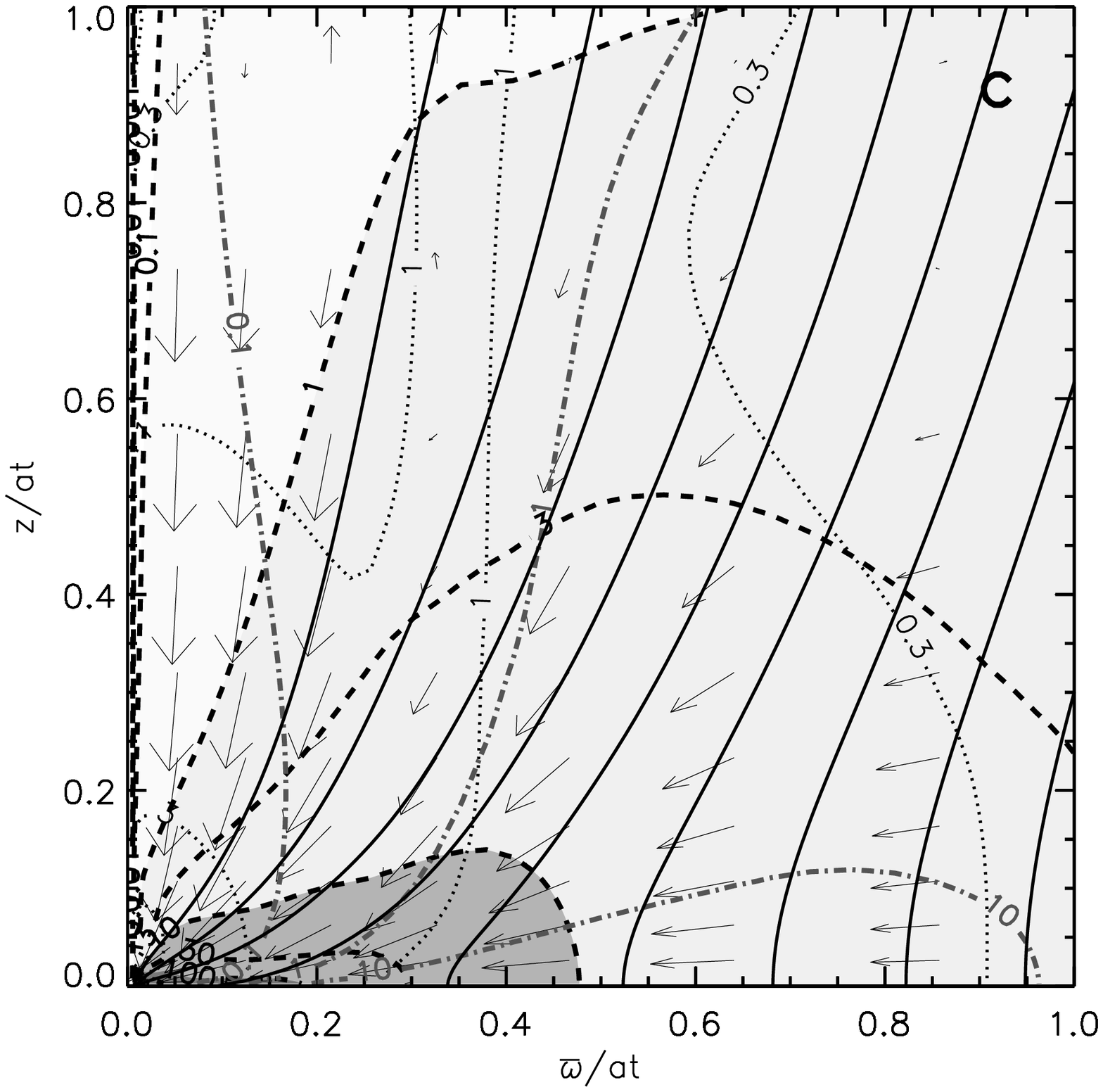} 
        \epsfxsize=.48\textwidth \epsfbox{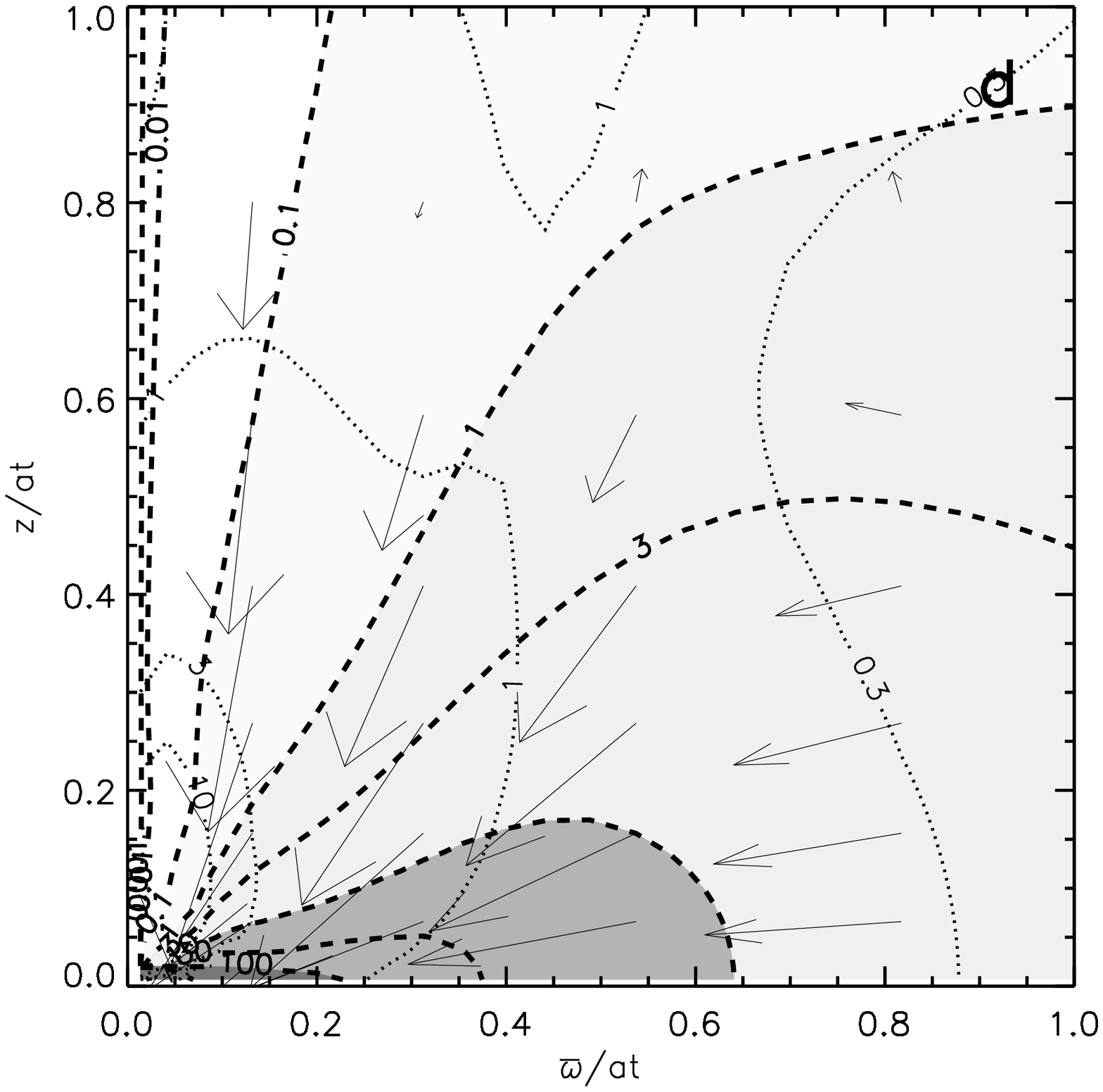}
\caption{Plots of self-similar density, $4\pi Gt^2\rho$, velocity field, 
${\bf u}/a$ and magnetic field, ${\bf B}G^{1/2}a/t$, lines for 
$v_0=0.25$ and different values of $H_0=0.03125, 
0.125, 0.25$ and $0.5$.  The isodensity 
contours are plotted as dashed lines, with the shades highlighting the
high density regions. The magnetic field lines are plotted as solid lines,  
with contours of constant $\beta$ (dash-dotted) superposed. The
velocity in every fifth cell is shown by unit vectors, with its magnitude 
given by the dotted contours. Field lines are not the same across all 
panels; the relative strength of magnetic field is indicated by the value 
of the plasma $\beta$.}
\label{fig:allss2}
\end{figure}

\begin{figure}
    \centering
    \leavevmode
        \epsfxsize=.48\textwidth \epsfbox{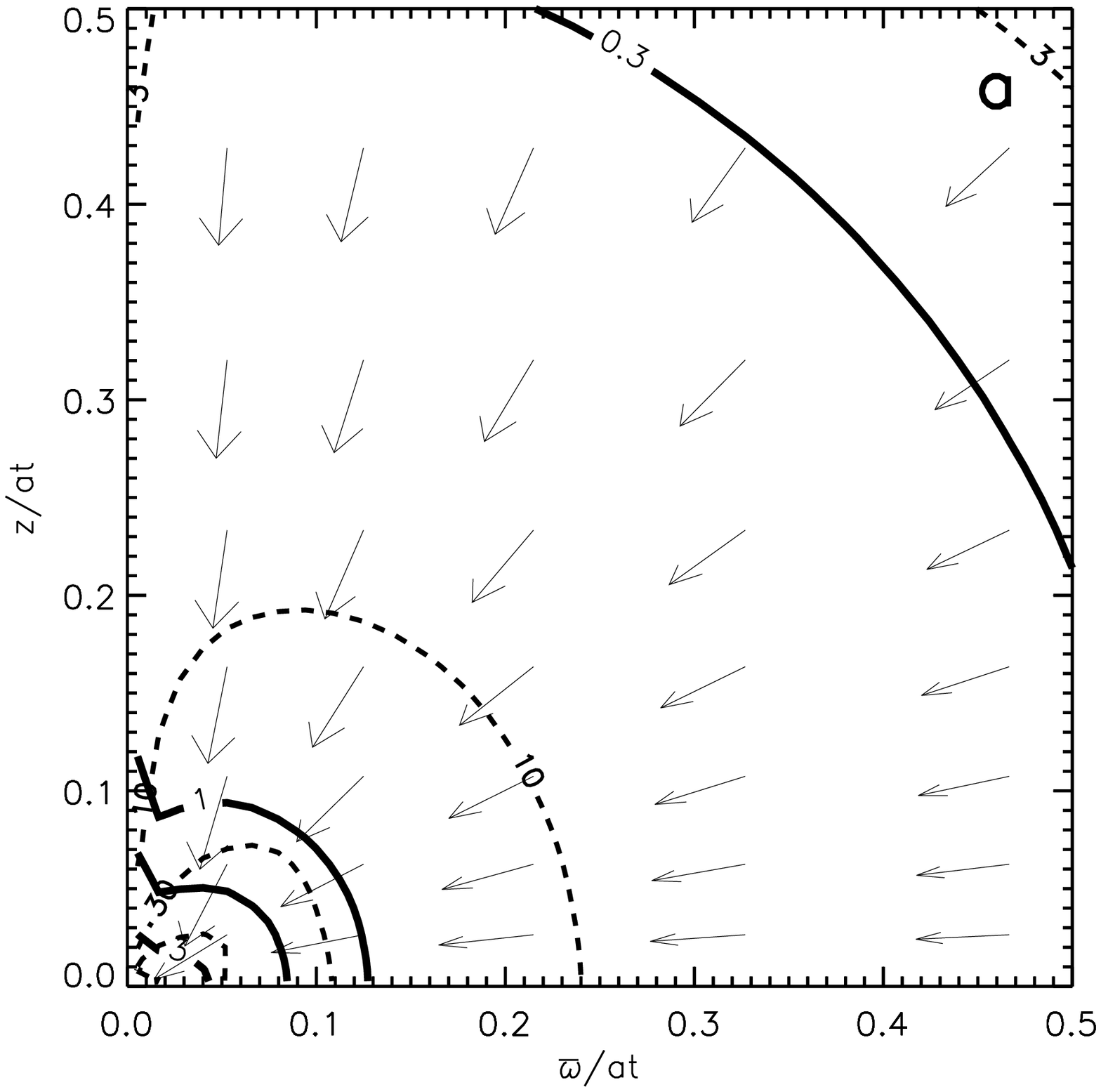} 
        \epsfxsize=.48\textwidth \epsfbox{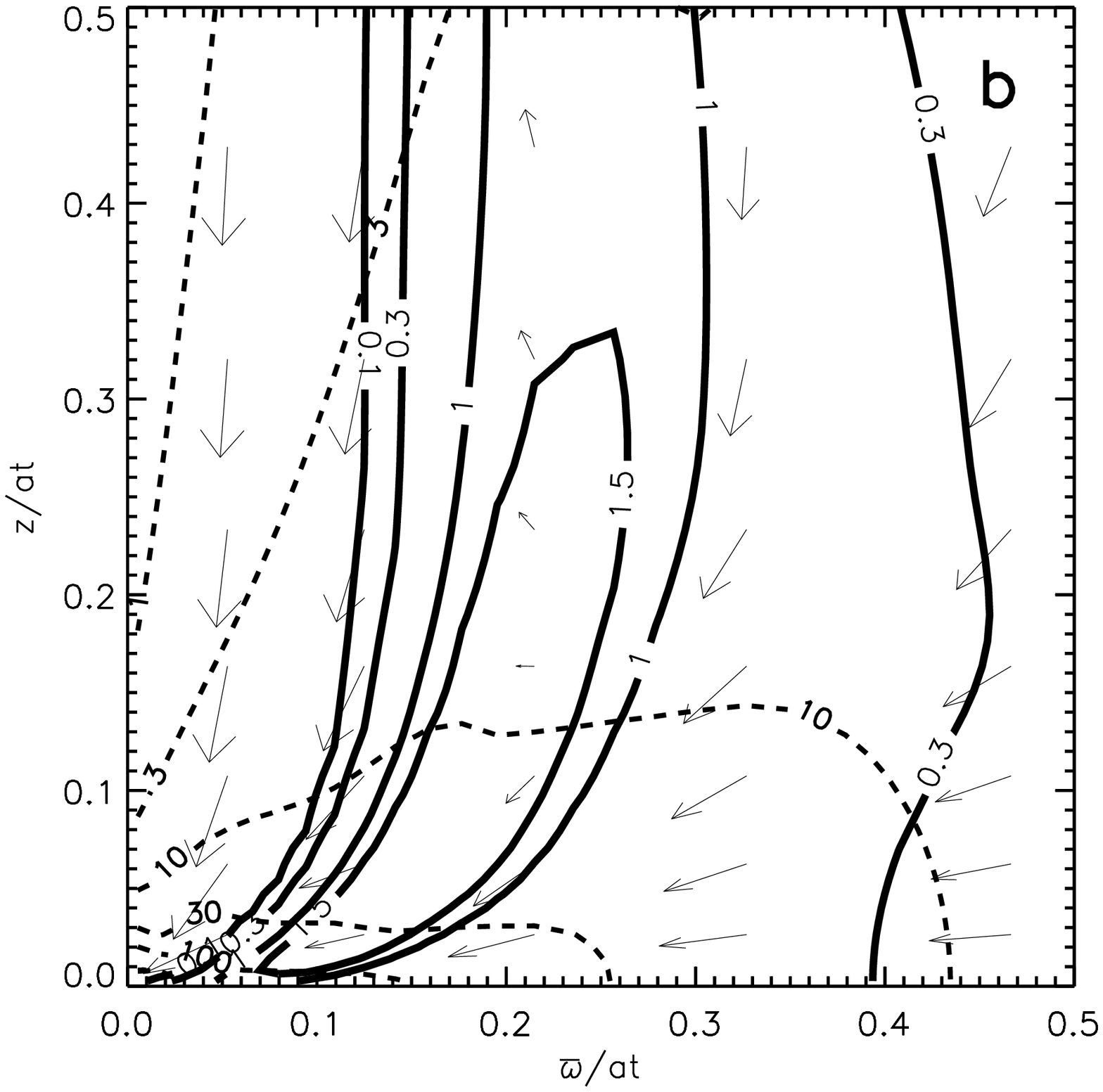} 
        \epsfxsize=.48\textwidth \epsfbox{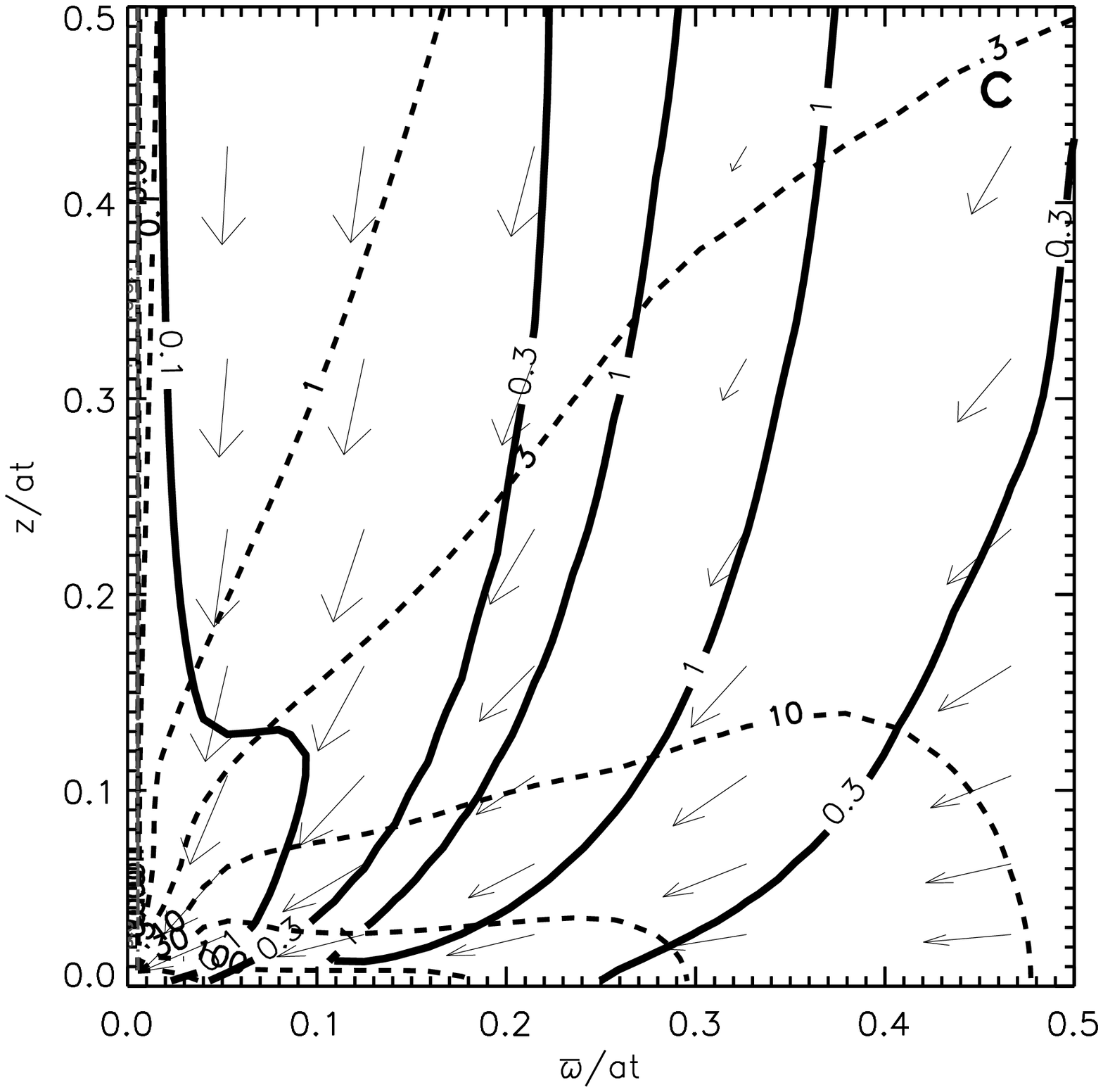} 
        \epsfxsize=.48\textwidth \epsfbox{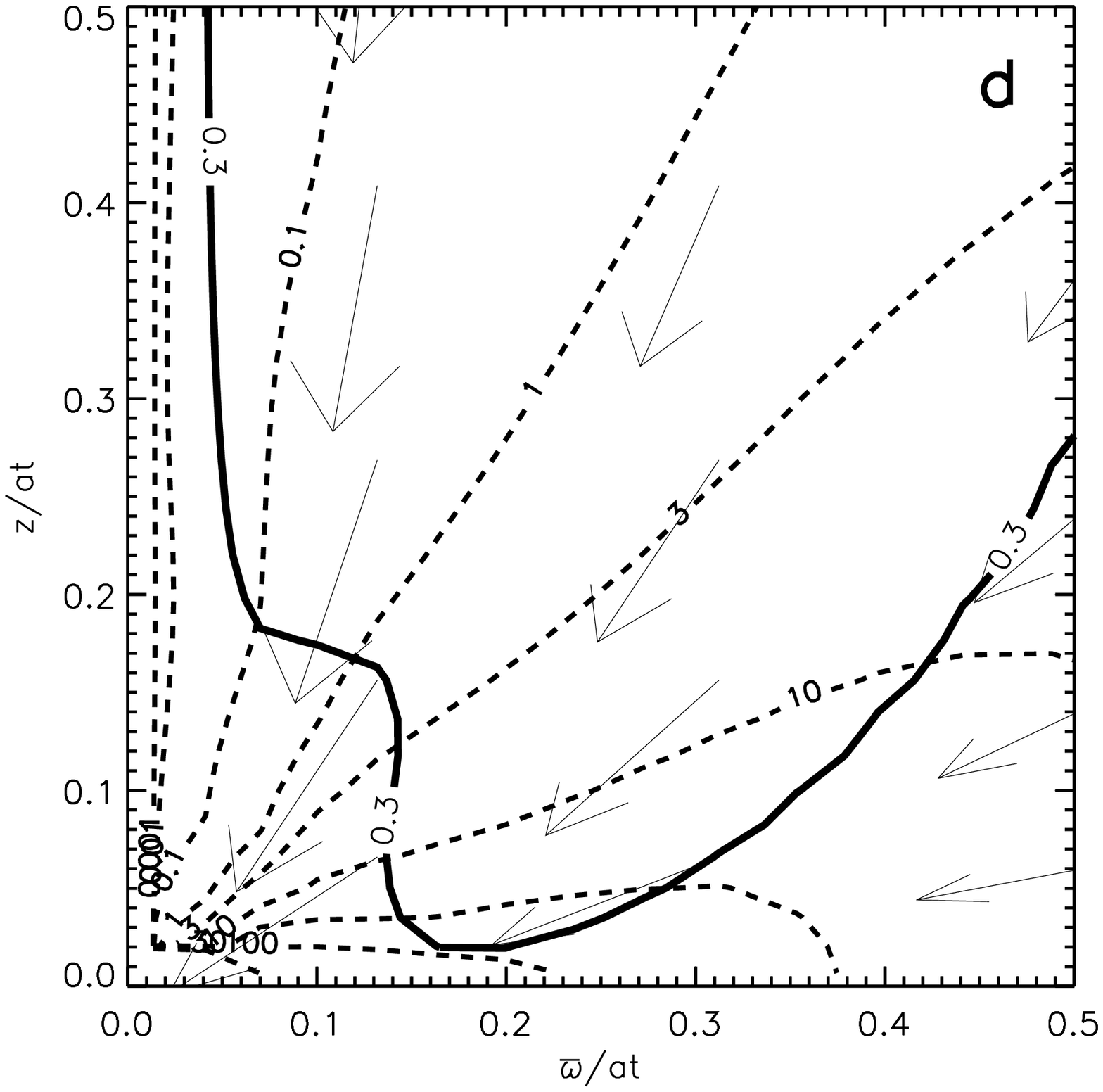}
\caption{Plots of self-similar density (dashed contours) and rotational 
speed (solid contours) for 
$v_0=0.25$ and different values of $H_0=0.03125, 0.125, 0.25$ 
and $0.5$. Comparing the rotational contours shows that the 
pseudodisk rotates more slowly in a more strongly magnetized case. 
}
\label{fig:ang}
\end{figure}

\end{document}